\RequirePackage{fix-cm}
\documentclass[twocolumn,epjc3]{svjour3}  
\smartqed  
\RequirePackage{graphicx}
\RequirePackage{amsmath}
\RequirePackage[T1]{fontenc}
\RequirePackage{xcolor}
\RequirePackage{lineno}
\RequirePackage{float}
\RequirePackage{dblfloatfix}
\RequirePackage{flushend}
\RequirePackage{array}
\newcolumntype{P}[1]{>{\raggedright\arraybackslash}p{#1}}

\RequirePackage[colorlinks,citecolor=blue,urlcolor=blue,linkcolor=blue]{hyperref}
\RequirePackage{etoolbox}

\makeatletter
\patchcmd{\equation}{\incr@eqnum\mathdisplay@push}{%
  \ifhmode
    \unskip
    \vadjust{%
      \incr@eqnum
      \global\let\gfix@incr@eqnum\incr@eqnum
      \global\let\gfix@currentlabel\@currentlabel
      \global\let\gfix@currentHref\@currentHref
    }%
    \let\incr@eqnum\gfix@incr@eqnum
    \let\@currentlabel\gfix@currentlabel
    \let\@currentHref\gfix@currentHref
  \else
    \incr@eqnum
  \fi
  \mathdisplay@push
}{}{}
\makeatother

\journalname{Eur. Phys. J. C}
\begin{document}

\title{Measurements of Angle-Resolved Reflectivity of PTFE in Liquid Xenon with IBEX 
}

\author{S.~Kravitz\thanksref{aLBNL,e1}
       \and R.~J.~Smith\thanksref{aUCB}
       \and L.~Hagaman\thanksref{aUCB}
       \and E.P.~Bernard\thanksref{aUCB}
       \and D.N.~McKinsey\thanksref{aLBNL,aUCB}
       \and L.~Rudd\thanksref{aUCB}
       \and L.~Tvrznikova\thanksref{aLBNL,aUCB,aYale,eLLNL}
       \and G.~D.~Orebi~Gann\thanksref{aLBNL,aUCB}
       \and M.~Sakai\thanksref{aLBNL,aUCB}
}

\institute{Lawrence Berkeley National Laboratory, 1 Cyclotron Rd., Berkeley, CA 94720, USA\label{aLBNL}
          \and
          University of California Berkeley, Department of Physics, Berkeley, CA 94720, USA\label{aUCB}
          \and
          Yale University, Department of Physics, 217 Prospect St., New Haven, CT 06511, USA\label{aYale}   
}

\thankstext{e1}{e-mail: swkravitz@lbl.gov}
\thankstext{eLLNL}{Now at Lawrence Livermore National Laboratory}

\date{} 

\maketitle

\begin{abstract}
Liquid xenon particle detectors rely on excellent light collection efficiency for their performance. This depends on the high reflectivity of polytetrafluoroethylene (PTFE) at the xenon scintillation wavelength of 178~nm, but the angular dependence of this reflectivity is not well-understood. IBEX is designed to directly measure the angular distribution of xenon scintillation light reflected off PTFE in liquid xenon. These measurements are fully described by a microphysical reflectivity model with few free parameters. Dependence on PTFE type, surface finish, xenon pressure, and wavelength of incident light is explored. Total internal reflection is observed, which results in the dominance of specular over diffuse reflection and a reflectivity near 100\% for high angles of incidence. 


\end{abstract}

\section{Introduction}\label{sec:intro}
Liquid xenon (LXe) particle detectors are used in a variety of experiments including searches for dark matter \cite{LUX-NIM,XENON1T-expt,PandaX-expt}, neutrinoless double beta decay \cite{EXO-expt}, and non-standard muon decay \cite{MEG-results}. These experiments detect energy deposits in the LXe through its scintillation light, hence critical detector parameters such as energy threshold and resolution depend strongly on the light collection efficiency. A high efficiency can be achieved by surrounding the active LXe volume with material that reflects well at the scintillation peak of 178~nm \cite{Xe-scint}. While typical construction materials such as stainless steel have poor reflectance in the UV, the polymer polytetrafluoroethylene (PTFE) is considerably more reflective.

Studies from LXe particle detectors suggest that the total reflectance of PTFE at LXe scintillation wavelengths is >90\% when immersed in LXe \cite{LUX-comprehensive,Yamashita-PTFE,Neves-IEEE}. These results are based on simulations assuming reflection is purely diffuse, and exhibit degeneracies with other parameters such as attenuation length in the LXe. A dedicated experiment designed to measure total PTFE reflectivity in LXe found values in excess of 95\% across several PTFE types \cite{Neves:2016}. Though some of these PTFE surfaces show signs of deviation from purely diffuse reflection, the experiment referenced is unable to directly measure the angular distribution of the reflected light.

More precise detector simulation requires a model of PTFE reflectivity as a function of both incident and outgoing angle - the bi-directional reflectance intensity distribution function (BRIDF). BRIDF measurements have been performed in vacuum at Xe scintillation wavelengths \cite{Silva:2009,Choi_thesis,Bokeloh_thesis}, indicating a significant specular component and much lower total reflectivity than observed in LXe. A model for PTFE BRIDF in vacuum, parametrized by physical quantities such as PTFE refractive index, albedo, and surface roughness, captures all of the observed features \cite{Silva:2010}. However, extrapolating this model to measurements in LXe is not straightforward, and attempts to do this still underestimate the observed total reflectivity \cite{Silva_thesis}. 

The setup described in \cite{Bokeloh_thesis} has also been used for angle-resolved measurements in LXe \cite{Levy_thesis,Kaminsky_thesis}. These also clearly demonstrate a significant specular component at high incident angles, as is qualitatively expected from Fresnel reflection in a medium of higher refractive index. However, these measurements suffer from systematic uncertainties such as angle calibration and incident beam normalization, which limit their quantitative power. Furthermore, though measurements in \cite{Kaminsky_thesis} were taken at discrete angles outside of the plane of incidence, the empirical models used to describe the data do not allow for extrapolation to the full hemisphere, and hence cannot provide the full reflectivity profile or the integrated hemispherical reflectance.

The Immersed BRIDF Experiment in Xenon (IBEX) is designed to directly measure the angular distribution of Xe scintillation light reflected off PTFE submerged in LXe. The data presented here, along with the microphysical model used to explain them, complement total reflectivity measurements and allow a complete simulation of light propagation in LXe detectors employing PTFE reflectors.

\section{Experimental Setup}\label{sec:setup}
\begin{figure}

  \centering\includegraphics[width=0.5\textwidth]{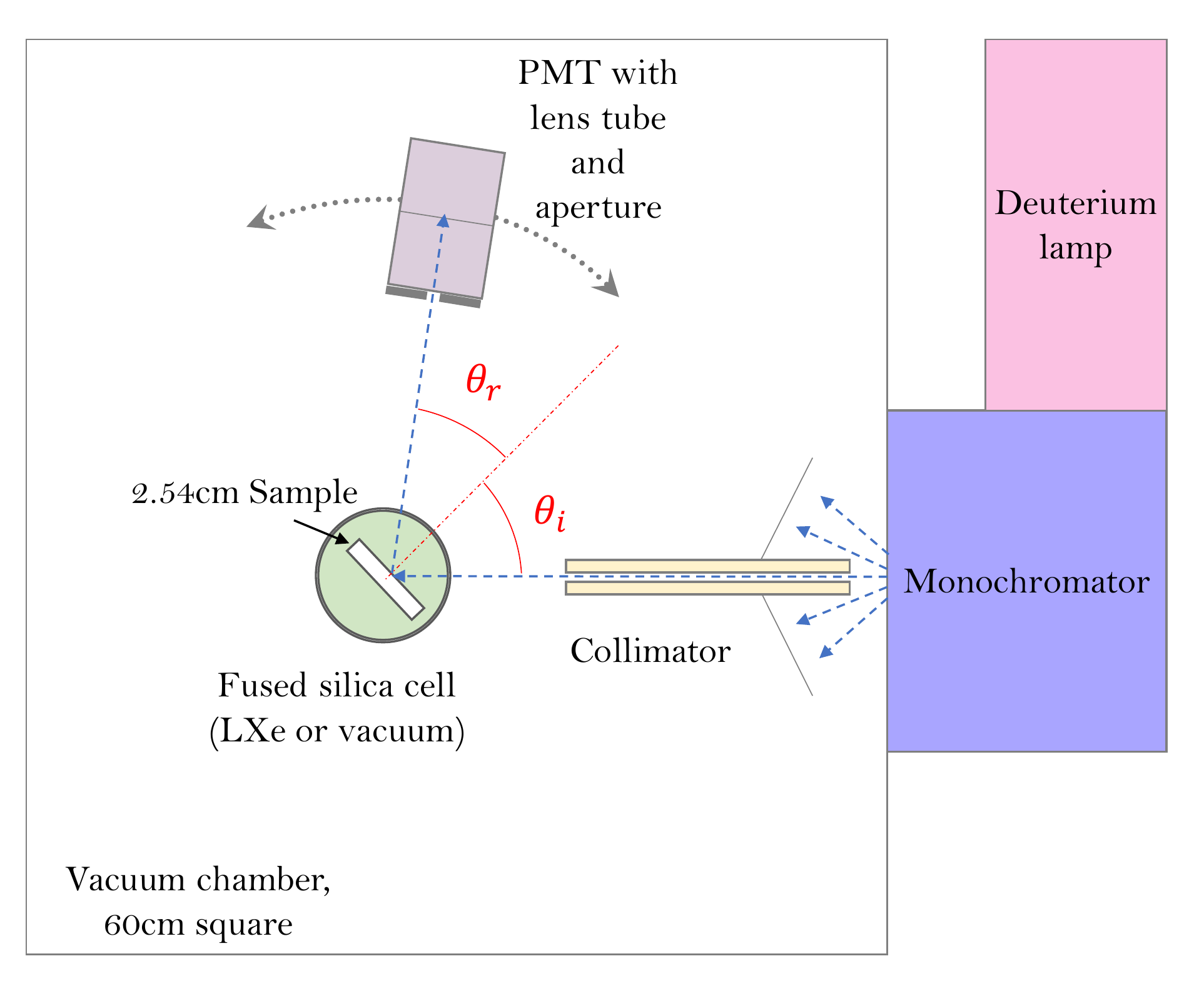}

\caption{Schematic of IBEX optical components as seen from above, not to scale. The fused silica cell can be evacuated or filled with xenon, with the sample at the center of the cell.}
\label{fig:IBEX-schematic} 
\end{figure}

The central components of IBEX are a fixed beam of light, an optical cell that may be filled with LXe or pumped to vacuum, a PTFE sample in the path of the beam and contained in the cell, and a photomultiplier tube (PMT), as shown in Fig.~\ref{fig:IBEX-schematic}. The beam is incident on the flat face of the PTFE sample. Measurements of surface roughness confirm the surface to have no significant features within the illuminated area. The origin of the coordinate system is defined to be the illuminated point on the sample. The angle between the vector normal to the sample surface and the vector in the reverse direction of the incident beam is called the incident angle $\theta_i$ and can be varied by rotation of the sample about an axis in the plane of its flat surface. Upon encountering the PTFE sample, a portion of the light is reflected. The PMT is used to measure the amount of reflected light, and can be rotated about the origin to view the sample from different angles.

Rays from the illuminated spot into the hemisphere into which reflection may occur are parametrized by two angles as follows: project the ray into the plane of incidence containing the sample normal and the beam direction. The viewing angle $\theta_r$ is the angle between the ray's projection and the sample normal, and $\phi_r$ is the angle between the ray and its projection. Reflection can occur into the full hemisphere of angles in which both $\theta_r$ and $\phi_r$ range between -90$^\circ$ and 90$^\circ$, but the PMT is allowed to rotate about the origin only in the plane of incidence, so reflected light is detected only in the region where $\phi_r$ is near 0. Measurements presented here are rates of reflected photons detected by the PMT at various values of $\theta_i$ and $\theta_r$, normalized by the solid angle of the PMT aperture as viewed from the origin and by the incident photon rate on the sample. These measurements are compared to models of BRIDFs, defined as
\begin{align}\label{eqn:BRIDF}
 \varrho(\theta_i,\theta_r,&\,\phi_r) =  \frac{d\Phi_r/d\Omega_r}{\Phi_i}
\end{align}
where $d\Phi_r$ is the intensity of the light reflected into an infinitesimal solid angle $d\Omega_r$ at the viewing position $(\theta_r,\phi_r)$, and $\Phi_i$ is the intensity of the incident light. This is exactly the normalized measured quantity described above in the idealized limit of infinitesimal beam width and PMT aperture.

\subsection{Apparatus}\label{sec:apparatus}
\begin{figure}

  \centering\includegraphics[width=0.4\textwidth]{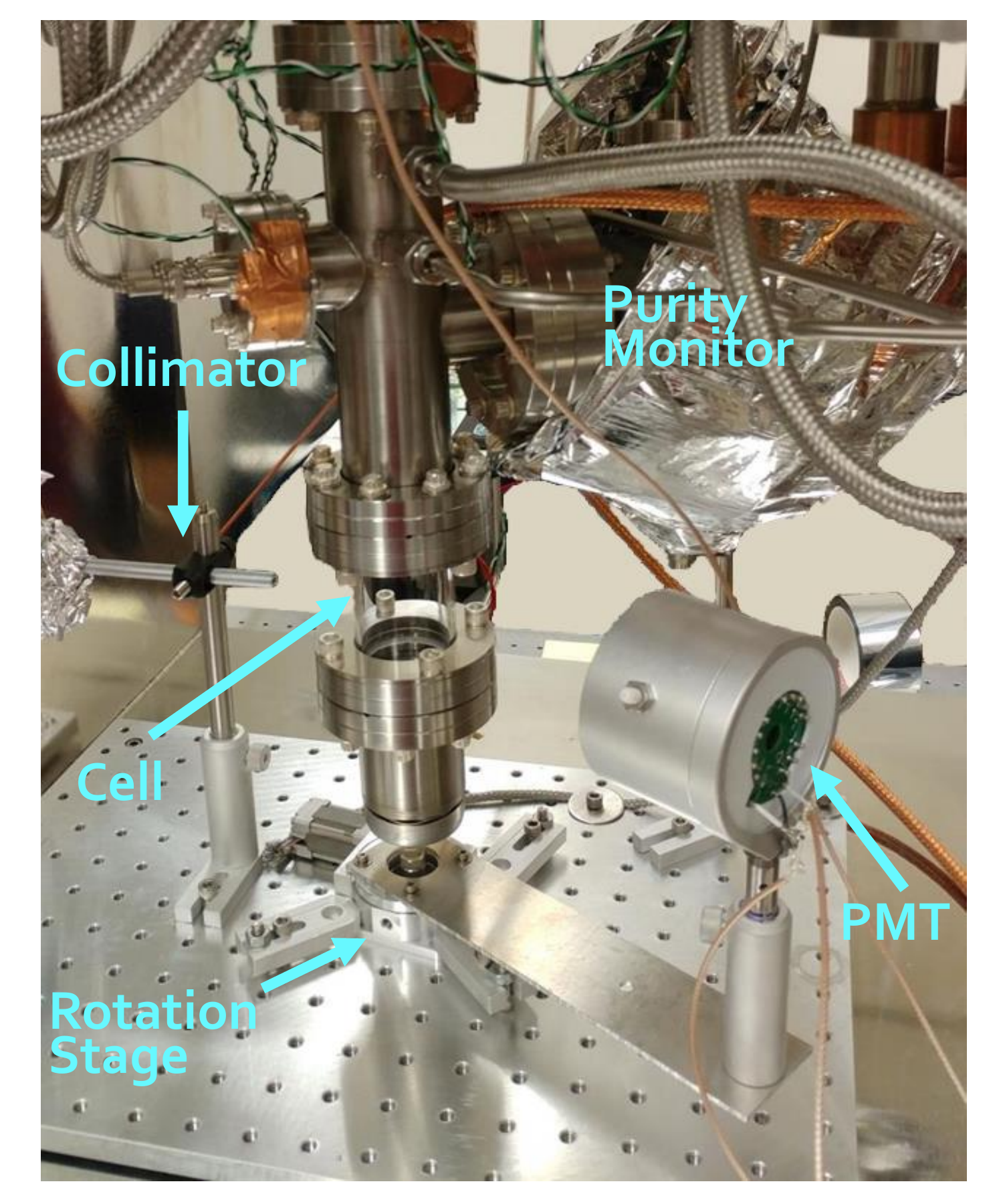}

\caption{Photo of IBEX before the addition of multi-layer insulation.}
\label{fig:IBEX-photo} 
\end{figure}

\begin{figure}

  \centering\includegraphics[width=0.4\textwidth]{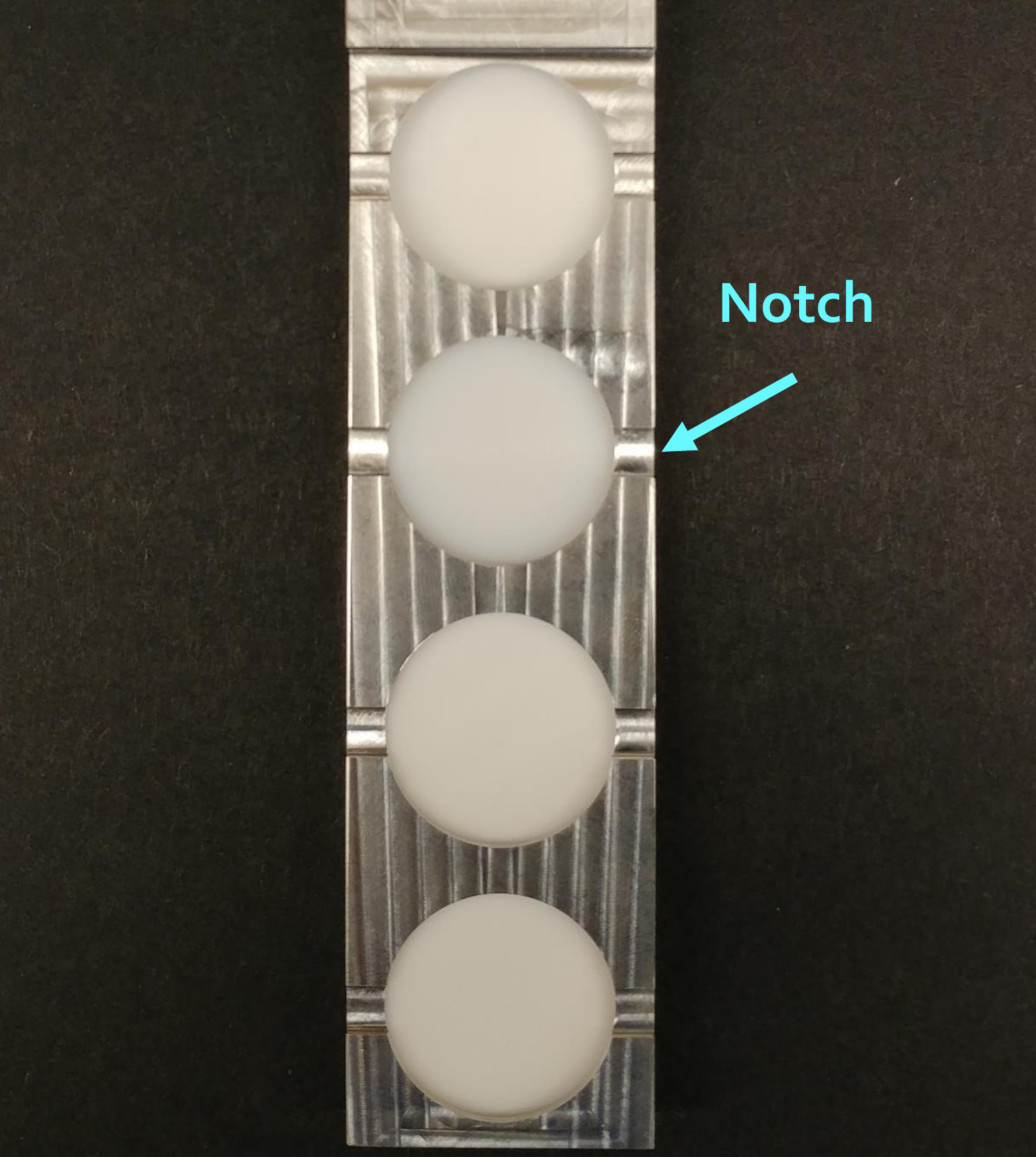}

\caption{Photo of sample rack with PTFE samples.}
\label{fig:sample-rack} 
\end{figure}

The setup broadly described above resides in a vacuum chamber on an optics table (Fig.~\ref{fig:IBEX-photo}). Inside the vacuum chamber, an independently sealed inner volume can be separately pumped out or filled with xenon via feedthroughs at the top of the vacuum chamber connecting to gas lines, which in turn connect to a gas panel. Pumping out the main vacuum chamber allows for operation of the deuterium lamp light source, transmission of light at vacuum ultraviolet (VUV) wavelengths, and thermal insulation of the inner volume. Typical pressures for the vacuum chamber during operation are in the $10^{-4}$~mbar range. The inner volume, aside from the fused silica cell, is covered with multi-layer insulation to reduce radiative heating of the xenon. A cylindrical column spans the center of the vacuum chamber and is mounted to a port at the top of the vacuum chamber. At the top of this port, a rotary-linear feedthrough connected sample transfer arm enables manual manipulation of the PTFE sample positions \textit{in situ}. Samples are held in a sample rack with four sample spaces arranged in a vertical array (Fig.~\ref{fig:sample-rack}). The sample transfer arm controls the vertical position of the sample rack to select which sample is at the height of the incident light, and has a dial at its top which is rotated to determine the incident angle. The section of the column at the height of the incident beam is a fused silica optical cell, selected for transmission of VUV light, with the axis of rotation of the sample rack coinciding with the symmetry axis of the cell. The PMT is mounted to an electronically controlled vacuum-compatible rotation stage which sets the viewing angle $\theta_r$ and is fixed to an optical breadboard at the floor of the vacuum chamber.  

The light source in IBEX consists of a deuterium lamp and monochromator coupled to one side of the vacuum chamber. Light is generated by a McPherson Model 632 deuterium lamp, which outputs a spectrum that is broad and continuous in the regime of UV wavelengths above 165~nm used here. This light is passed to the McPherson 234/302 monochromator with a 1200~g/mm diffraction grating, which is capable of selecting wavelengths below 550~nm. The efficiency of the grating differs by <3\% at 178~nm for the two polarization modes, leaving the outgoing beam unpolarized. An exit slit at the monochromator can be adjusted in size during operation to vary the beam intensity. This system is specified to have a wavelength dispersion of 4~nm in full width at half maximum per mm of width at the exit slit. This was verified and the spectral shape determined to be roughly Gaussian using a spectrometer for wavelengths above 200~nm \cite{Benson2018}. The exit slit is adjusted to compensate for changes both in lamp output when measuring at wavelengths other than 178~nm as well as in LXe purity. The maximum slit size used was 1.15~mm, corresponding to <2~nm in wavelength resolution. Slit size testing indicated no significant change in the beam spatial profile across this range. The monochromator is set to select 178~nm light, which then enters the main vacuum space. An aluminum pipe of length 178~mm and inner diameter 3.56~mm is mounted on the optical breadboard in the vacuum chamber and placed in the path of the incoming light to collimate it into a narrow beam, which is aimed at normal incidence to the curved face of the cylindrical fused silica cell.

Significant emphasis in the design phase of IBEX was placed on the fused silica cell. It is made of Suprasil 310, a fused silica material which, according to specifications, transmits close to 90\% of 178~nm light at normal incidence with a thickness of 2~mm, including both absorption in the material and reflection at the boundaries. It has a height of 76~mm, with inner diameter 42~mm and outer diameter 48~mm. Spring-loaded PTFE o-rings are used to make the quartz-metal seals at both ends of the cell. This design avoids introduction of
bolts crossing the length of the cell that would obstruct
optical measurements. A spring couples the bottom of the stainless steel column below the cell to the optical breadboard, and can be tightened to hold the assembly in place even when the column length varies due to thermal contraction. The incident light is normal to the cell's surface and the illuminated spot on the sample axis is centered in the cylindrical cell to minimize effects of refraction at the cell boundaries either prior to or after reflection from the sample.

Light is detected by a Hamamatsu R6041-06 PMT, sensitive in the wavelength range 160~nm to 650~nm. An aluminum housing surrounds the PMT, with an aperture restricting the light-sensitive area at its front to improve resolution on the viewing angle and to reduce backgrounds from stray light. The aperture is a circle with adjustable diameter set to 9.8~mm for sample measurements, corresponding to a width of 4 degrees in viewing angle from the PMT rotation axis. The PMT remains at room temperature during both vacuum and LXe measurements.

Single photon pulse counting is performed on the PMT signal by a series of NIM modules. First the signal is amplified, then sent to a discriminator, which produces a square pulse when the amplitude of the input signal exceeds a threshold. The discriminator output is amplified and fed to a ratemeter. The ratemeter outputs a voltage logarithmically proportional to the pulse count rate; this output is digitized and converted to a rate measurement. At each PMT position, 50 samples of the ratemeter measurement are taken, from which the median and standard deviation are recorded. Raw digitized waveforms from the PMT and discriminator outputs were analyzed to optimize the discriminator threshold and to validate the linearity of the readout scheme over pulse rates ranging from $\sim$100~Hz (the scale of the typical PMT dark rate) to $\sim$MHz (the scale of the maximum rates observed in measurements of the incident intensity). Linearity was further verified by demonstrating that reflectivity measurements of a sample in vacuum at 100~kHz and 1~MHz incident photon rates agreed within uncertainty.

Gaseous xenon (GXe) is delivered to the apparatus via a gas panel, which includes a circulation pump, heated zirconium getter, and additional chemical purifiers. The inlet GXe tube is fed through the vacuum chamber to a heat exchanger coupled to a pulse tube refrigerator (PTR). GXe condenses to LXe in the outer volume of the heat exchanger, and drips to the bottom of the central column below the fused silica cell. As LXe continues to condense, the liquid level fills the cell and eventually exceeds the height of a binary capacitive level gauge above the cell. At this point the LXe fill is considered complete. It is also possible to view the cell directly through a viewport on the side of the chamber, from which one can observe the liquid level if it is in the cell or determine if the LXe is boiling or turbulent. During measurements, the viewport is sealed with a light-tight cap. The top of the column is connected to the inner volume of the heat exchanger, from which GXe is fed to an outlet at the top of the vacuum chamber to recirculate through the gas panel. The gas panel is described in greater detail in \cite{xebra}. A Xe purity monitor, the design of which is also described further in \cite{xebra}, is attached to the column. However, due to failure of the getter, purity was too poor to be measurable with the purity monitor during this study. Relative purity can instead be judged by transmission of UV light through the LXe, and is observed not to have a significant effect on measurements (Sec.~\ref{sec:systematics}).

Conditions of the xenon and vacuum spaces are monitored and manipulated with a programmable logic control system. GXe pressure is measured at the vacuum chamber outlet as well as at various spots on the gas panel. To convert from the measured GXe pressure at the vacuum chamber outlet, above the surface of the LXe, to the LXe pressure at the sample location, 0.05~bar of additional pressure due to the column of LXe above the sample is added. It is this inferred LXe pressure which is reported here alongside BRIDF measurements in LXe. Resistance temperature detectors measure the temperature at the PTR, the inner and outer volumes of the heat exchanger, and on the column below and above the fused silica cell and at the level gauge location. The pressure of the vacuum chamber is also monitored, as well as the capacitance of the level gauge. To achieve or maintain desired conditions, the user may adjust the power of heaters attached to the PTR and heat exchanger, and the circulation pump speed.

\subsection{Alignment}\label{sec:alignment}
To ensure that the measurements taken in this setup closely match the idealized geometry described above, a number of alignment procedures are completed, described in detail here and summarized in the appendix in Tab.~\ref{tab:alignments}. Several components of the setup are fixed in place permanently and need be aligned only once prior to the measurement campaign. The beam must be incident on the samples along the axis of rotation; the light from the monochromator must pass through the collimator; and the PMT rotation axis, sample rotation axis, and symmetry axis of the cell must all be coincident. Each of these alignments have been tested and shown to be stable between and during measurements.

Since the collimator determines the alignment of the beam, aiming the beam at the sample rotation axis can be accomplished regardless of light source. Using a laser instead of the light from the deuterium lamp facilitates the alignment as it does not require the slow process of pumping out the chamber between adjustments of the collimator position. To monitor the beam alignment to the sample rotation axis, a rectangular alignment sample with height equal to the 2.54~cm sample diameter, but with much narrower width, is fabricated from PTFE. It is placed vertically in the sample rack, spanning the sample slot along its axis of rotation, and preliminary alignment is achieved by adjusting the collimator position until an illuminated spot is visible on the alignment sample. The alignment was further refined by observing symmetry of the alignment sample's shadow when scanning the PMT behind it.

Once the collimator was aligned to the rotation axis of the samples, the laser was removed and the monochromator was aligned to the collimator. This alignment need only suffice to achieve significant light transmission through the collimator. Measurements of the photon rate in the beam through the collimator were taken following slight adjustments of the monochromator orientation and position until it appeared that the photon rate is near maximal. The beam is also sufficiently powerful at visible wavelengths to verify by eye that light is passing through the monochromator, using the viewport at the opposite end of the chamber. Since the monochromator is coupled to the vacuum chamber with a flexible bellows, it can be repositioned while the chamber is at vacuum. After alignment of the monochromator, it was anchored firmly to the optics table.

The rotation stage is mounted to the optical breadboard inside the vacuum chamber. The breadboard offers both precise manipulation of the rotation stage as well as a means to securely fix its position after alignment. The rotation stage has a bore at its center which is similar in diameter to the width of the sample rack, so by lowering the sample rack into the bore, a satisfactory preliminary alignment between the PMT rotation axis and sample rotation axis can be completed by eye. This alignment is tested more precisely with mirror measurements as discussed below. Mirror measurements during the first LXe run and afterwards in vacuum indicated misalignment of the rotation stage, likely due to mistakenly bumping or otherwise shifting it. Based on those measurements, a displacement in the rotation stage position by 2.5~mm was inferred and corrected prior to the second LXe run. This displacement resulted in a discrepancy between the viewing angle coordinate $\theta_r$ described above and the actual angle of the rotation stage. The mapping between the desired coordinate system and the actual angles due to this misalignment includes a constant shift at zero order. Such shift is corrected by shifting the incident angle used in the fits, which is equivalent to a constant shift in viewing angle. For this data, shifts of $\sim3^\circ$ in incident angle relative to nominal values were typical. Of the data reported here, this misalignment only affected the M18 sample in LXe and the LUX sample in both LXe and vacuum.

In order to avoid systematics due to refraction and reflection at the boundaries of the fused silica cell, it is desirable for the symmetry axis of the cylindrical cell to be coincident with the sample rotation axis to which the beam is aligned. The cell and the column it is attached to can be tilted from the top of the chamber by adjustment of a bellows. Since the sample rack width is close to the inner diameter of the flange at the top of the cell, smooth rotation is only possible when the sample rotation axis and cell symmetry axis are closely aligned, and is therefore the benchmark of satisfactory positioning of the cell. Further precision is difficult to maintain as minor changes in the column orientation occur due to thermal contraction when it is filled with LXe. Misalignment of the beam with the cell results in shifts of $\sim4^\circ$ in the beam position in LXe relative to vacuum during power measurements, which are described in Sec.~\ref{sec:power}.

Additional alignments that are relatively stable across measurements, but were repeated at least once during the course of the experiment, are the alignments of the sample and PMT heights to the plane of incidence.

In order to maximize the portion of the beam that strikes the sample, the sample height is set so that the beam is incident on its equator. The alignment to the equator of the sample relies on a feature of the sample rack. The samples are recessed into the sample rack, but near each sample's equator, a notch is cut into the face of the sample rack on either side of the sample, as visible in Fig.~\ref{fig:sample-rack}. This notch prevents obstruction of the sample by the sample rack at values of $\theta_i$ and $\theta_r$ far from the sample normal. By holding the sample at high incident angle, a significant upwards step in intensity is observed when the height of the notch matches the beam height. After locating this height, a marking is made on the sample transfer arm on the side exposed to atmosphere. This procedure is repeated for each sample, resulting in a set of four markings indicating the height to which the sample transfer arm should be set during sample measurements.

Analysis of the measurements described here assumes that the PMT rotates in the plane of incidence, i.e. $\phi_r=0$. To set the PMT height so that it is in this plane, the aperture is narrowed to 2.3~mm in diameter, the sample rack is lifted clear of the cell, and the PMT is scanned about the location of the transmitted beam. The PMT height is adjusted to the location where the maximal incident photon rate is observed. These measurements characterize the beam profile as well as establishing that the PMT is within the plane of incidence at this angle. In vacuum, the beam profile is well-modelled as a radially symmetric Gaussian with a width of $\sigma = 0.9^\circ$ with respect to the PMT rotation axis. In LXe, the beam profile versus PMT height is not measured, as PMT height adjustment requires breaking vacuum and hence Xe recovery. Instead, power measurements in LXe as described in Sec.~\ref{sec:power} are compared with a model convolving the PMT aperture with a Gaussian of differing horizontal and vertical widths. These measurements are consistent with the vertical width of the beam being unchanged while the horizontal width is increased by a factor of 1.7, a magnification expected due to lensing at the interface with the cylindrical cell. To verify that the PMT remains in the plane of incidence when rotated to other angles, measurements of the rate of photons reflected off a mirror in the sample rack are taken and shown to be consistent at multiple incident angles.

\subsection{Power measurements}\label{sec:power}
Once the setup is properly aligned, it remains to calibrate the incident and viewing angles, determine the incident beam power on samples, and correct for stray light. The corresponding procedures are described in these remaining subsections of Sec.~\ref{sec:setup} and summarized in the appendix in Tab.~\ref{tab:calibrations}. Power measurements of the unobstructed beam are taken regularly, including each time the deuterium lamp is turned on or the wavelength selected by the monochromator is changed. For such measurements, the PMT aperture is opened to the diameter of 9.8~mm used for sample measurements, and the PMT is swept about the incident beam in increments of 0.25$^\circ$ behind the quartz cell with the sample rack lifted clear above the beam. The maximum rate observed by the PMT is extracted from this measurement. Since the angular size of the samples when viewed from the light source is larger than that of the PMT aperture, it is possible for more light to be incident on the sample during a measurement than is captured by the PMT at any single position during a power measurement. The maximum rate observed during a power measurement is divided by a correction factor to account for this light missing the PMT aperture. The correction factor is equal to the proportion of the integral of the beam profile models described in Sec.~\ref{sec:alignment} that falls within the aperture. The maximum rate divided by this correction is equal to the incident light intensity $\Phi_i$ in Eq. \ref{eqn:BRIDF}. In addition to providing the incident rate for normalization of reflectance measurements, the position of the maximal beam intensity is used to define the origin of the absolute angular positions on the rotation stage. If the beam is not at normal incidence to the cylindrical cell, filling the cell with LXe alters the lensing of the beam and changes the beam position observed in a power measurement. Therefore these measurements are also useful in checking the alignment of the beam to the cell.

\subsection{Mirror measurements}\label{sec:mirror}
Reflectance measurements using a UV-enhanced aluminum mirror, designed for use at 250-450~nm, are also regularly applied to characterize the setup and calibrate the sample incident angles, including each time the sample rack has been removed and reinserted. This is done by determining the dial setting at which the reflected beam appears at the expected specular viewing angle. These measurements provide a powerful check against any hard-to-model effects such as refraction in the LXe. Such calibrations indicate that the uncertainty in incident angle achieved by reading the dial at the top of the sample transfer arm is $\sim$2$^\circ$. Therefore in our analysis, each incident angle is allowed to float within $\pm2^\circ$ in the fits. The effect of this uncertainty on the fits is explored further in Sec.~\ref{sec:systematics}.  

Mirror measurements are also used to correct the observed incident rate: at high incident angles, part of the beam may miss the sample face, particularly in the case of any misalignment between the beam and the sample axis. To account for this effect, observed in a minority of datasets, the ratio of the mirror-reflected peak rate at a given incident angle to that at low incident angles (where the full beam hits the sample and no dependence of the rate on angle is seen) is calculated. The ratio versus incident angle is linearly interpolated between the measured angles, and the incident rate used to normalize a sample measurement at a given incident angle is divided by the corresponding ratio. Discrepancies of less than 5\% with the maximal rate in a set of mirror measurements are ignored to avoid adding noise to the incident rates. 

For calibration of incident angle, mirror measurements are performed at the wavelength of interest (typically 178~nm); the mirror reflectance appears specular at all observed wavelengths. For the correction of incident rates, mirror measurements are taken at 400~nm, since the total reflectivity of the mirror is degraded at lower wavelengths.

\subsection{Background measurements}\label{sec:background}
 Regular background measurements are used to subtract out the effect of PMT dark count rate, light leaks into the vacuum chamber, and light scattered at the cell interface or in the LXe. To determine the background due to light leaks and PMT dark rate, the beam is turned off and the PMT is scanned through the full angle range in which reflectivity measurements are taken. For each reflectivity dataset, the dark background rate at each viewing angle is subtracted from the measured count rate at that angle. Additionally, beam background measurements are collected with the beam transmitted through the cell with no sample in place, again scanning the PMT through the full angle range in which reflectivity measurements are taken. The beam background measurement versus PMT angle is then scaled by the relative beam intensity between the (dark rate-subtracted) beam background measurement and the dataset and subtracted from the measured count rate at the same PMT position: 
\begin{equation} \label{eqn:bkg-corr}
R_c(\theta_r)=R_m(\theta_r)-R_d-\frac{\Phi_m}{\Phi_b}(R_b(\theta_r)-R_d),
\end{equation}
\noindent where $R(\theta_r)$ is the PMT count rate versus viewing angle for the corrected reflectivity measurement ($R_c$), uncorrected measurement ($R_m$), background ($R_b$), or dark counts ($R_d$) and $\Phi$ is the power measurement from the beam with no sample in place for the relevant dataset (Sec.~\ref{sec:power}). The power scaling factor $\frac{\Phi_m}{\Phi_b}$ differs from unity by <5\% for all cases except for measurements at wavelengths other than 178~nm (<12\%) and measurements from the last run in LXe (Sec.~\ref{sec:conditions}) which had a noticeable change in LXe purity and hence UV transmission during the run (<30\%). In the case of vacuum reflectivity measurements, the dark background rate dominates and the correction due to the beam background is ignored.

The correction in Eq.~\ref{eqn:bkg-corr} does not account for the change in scattered light caused by introducing the sample rack and sample into the beam path. However, for most cases, the background rates measured have a minor effect on the fits, which is quantified further in Sec.~\ref{sec:systematics}.

\section{Reflectivity Model}\label{sec:model}
The microphysical model used in this work for light reflected at the PTFE surface is based on \cite{Silva:2009,Silva_thesis}, which demonstrate that the model provides a faithful description of PTFE reflectivity in vacuum at UV wavelengths. A brief explanation of this model follows. In general, the total BRIDF $\varrho$ can be broken into its diffuse and specular components, $\varrho_D$ and $\varrho_S$, respectively. A complete description of the angular form of these components depends on the microstructure of the surface and inhomogeneities of the bulk material, which are not practical to map in detail. This necessitates the parameterization of the material properties in aggregate, which suffices when the material structure is adequately homogeneous over the region illuminated by the incident light.

\begin{figure}

  \centering\includegraphics[width=0.5\textwidth]{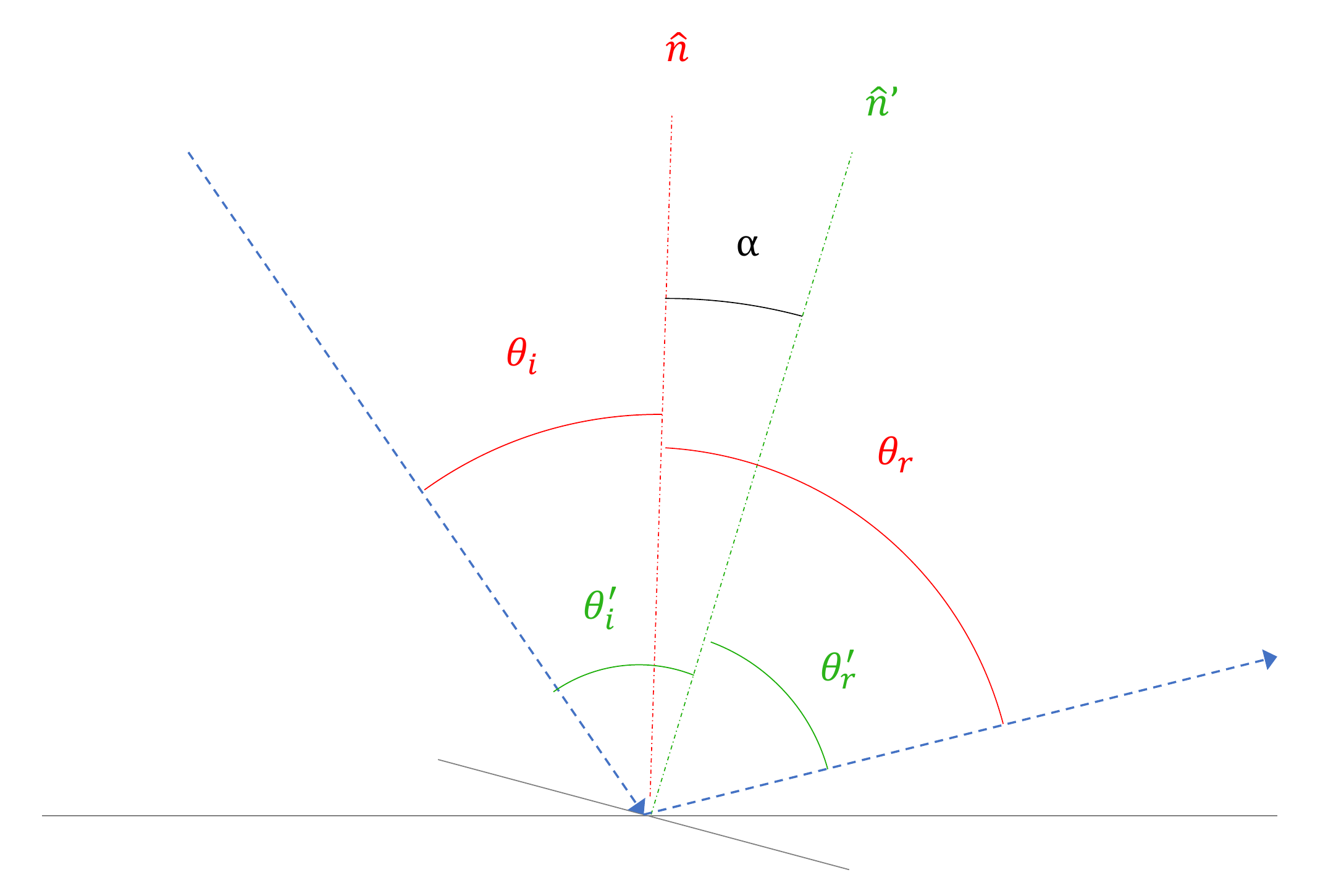}

\caption{System of angles and directions. The angle $\alpha$ is defined between the macrosurface normal $\mathbf{\hat{n}}$ and the local microfacet normal $\mathbf{\hat{n}'}$. Primed angles are defined relative to $\mathbf{\hat{n}'}$.}
\label{fig:local-angles} 
\end{figure}

The approach taken here models the surface as a series of microfacets with surface normals $\mathbf{\hat{n}'}$ randomly oriented according to a probability distribution function $P\left(\alpha\right)$, where $\alpha$ is the angle between the macrosurface normal $\mathbf{\hat{n}}$ and $\mathbf{\hat{n}'}$ (Fig.~\ref{fig:local-angles}). Primed angles ($\theta'_i$, $\theta'_r$) are defined relative to $\mathbf{\hat{n}'}$ rather than $\mathbf{\hat{n}}$. It is assumed that the arrangement of microfacets is isotropic, so that $P$ is independent of the azimuthal angle, $\phi_\alpha$, of the microfacet. The case of surfaces with a preferred direction can be handled by including an explicit $\phi_\alpha$ dependence for $P$ \cite{Ward_1992}. The distribution assumed here, originally formulated by Trowbridge and Reitz \cite{Trowbridge:75} (though sometimes referred to as the GGX model \cite{Walter:2007} in a computer graphics context), is given by:

\begin{equation}\label{eqn:P-TR}
 P(\alpha; \gamma) = \frac{\gamma^2}{\pi \cos^4\alpha\left(\gamma^2 + \tan^2\alpha\right)^2},
\end{equation}
\noindent where $\gamma$ parameterizes the surface roughness, and takes strictly positive values, typically below 1. A larger $\gamma$ describes a rougher surface and hence a wider distribution of microfacet angles. Alternative microfacet angle distributions, such as that from Beckmann \cite{Beckmann} (further refined by Cook and Torrance \cite{Cook-Torrance}), were considered but gave a poorer description of preliminary reflectivity measurements in vacuum in this work. 

Following this model, the specular component is given by:
\begin{align}\label{eqn:rho_s}
 \varrho_S(\theta_i,\theta_r,&\,\phi_r; n_0/n,\gamma) \nonumber  \\
 & = \frac{F(\theta_i',n_0/n) \cdot G(\theta_i,\theta_r,\phi_r; \gamma) \cdot P(\alpha_s; \gamma)}{4 \cos\theta_i}. 
\end{align}
Here $P(\alpha_s;\gamma)$ is evaluated at the microfacet angle $\alpha_s$ required for light incident at angle $\theta_i$ to be specularly reflected into viewing angle $\theta_r$. The term $F(\theta_i',n_0/n)$ is the Fresnel coefficient for reflection off PTFE of index $n$ submerged in a medium of index $n_0$, while $\theta_i'$ is the incident angle relative to the microfacet at angle $\alpha_s$. The shadowing and masking factor, $G(\theta_i,\theta_r,\phi_r; \gamma)$, accounts for light that is blocked from reflection at high incident or viewing angles due to microfacet protrusion. It is calculated assuming the Trowbridge-Reitz microfacet distribution (Eq.~\ref{eqn:P-TR}) according to the shadowing theory approximation by Smith \cite{Smith:1967}, carried out in \cite{Walter:2007}:%
\begin{align}
    G(\theta_i,\theta_r,&\,\phi_r ; \gamma) \nonumber \\
    & =  \Theta\left(\frac{\pi}{2}-\theta'_i\right)\Theta\left(\frac{\pi}{2}-\theta'_r\right) G'(\theta_i) G'(\theta_r)
\end{align}%
\begin{equation}
    G'(\theta; \gamma) \equiv \frac{2}{\left(1+\sqrt{1+\gamma^2\tan^2\theta}\right)}
\end{equation}
\noindent Here $\Theta$ is the Heaviside step function. The geometrical factor $1/(4 \cos\theta_i)$ in Eq.~\ref{eqn:rho_s} ensures that the integral $\int P(\alpha) \sin\theta_r d\theta_r d\phi_r$ is properly normalized, i.e. all light is reflected into one viewing angle or another in the limiting case $F=G=1$. A more precise calculation includes $G$ in the integral to be normalized. However, that integral has no analytic solution and this modification has little effect on the overall shape of $\varrho_S$ due to $G$ being very close to $1$ at most angles, therefore the approximation in Eq.~\ref{eqn:rho_s} is used.

In the case of a particularly smooth surface, Eq.~\ref{eqn:rho_s} may need to be modified to include a contribution from the coherent reflection, or specular spike. Kirchoff theory can be used to calculate the form of the coherent reflection in the case of conductive boundary conditions \cite{Beckmann}. The limits of this approach are explored in \cite{Ogilvy_1987}, and a modified empirical form that works well for PTFE in vacuum is found in \cite{Silva:2010}. With this addition, the specular component of the BRIDF has the form:
\begin{align}\label{eqn:rho_s_full}
 &\,\varrho_S(\theta_i,\theta_r,\phi_r; n_0/n,\gamma,K)  \nonumber  \\
  &\,= (1-\Lambda )\frac{F \cdot G \cdot P}{4 \cos\theta_i} + \Lambda \cdot F \cdot G \cdot \delta\left(\theta_i-\theta_r\right) \cdot \delta(\phi_r), 
\end{align}
where $\Lambda(\theta_i, \theta_r;K) = \exp\left( -K (\cos\theta_i+\cos\theta_r)/2 \right)$ and the delta functions ensure that the incident and viewing angles are the same, as required for specular reflection. Note that in the limit $K \to \infty$, $\Lambda$ vanishes and we recover the original specular component \eqref{eqn:rho_s}. This specular spike component is included when the fit to data is significantly improved by doing so, as is seen for the smoother samples considered here.

The diffuse component, which comes from sub-surface scattering of light by bulk inhomogeneities, is given by:
\begin{align}\label{eqn:rho_d}
\varrho_D(\theta_i,&\,\theta_r,\phi_r;\rho_l,n_0/n,\gamma) \nonumber \\ 
& = \frac{\rho_l}{\pi} N\left(\theta_i,\theta_r,\phi_r;\gamma\right) \cdot W(\theta_i,\theta_r;n_0/n) \cos\theta_r.
\end{align}
Here $\rho_l$ is the diffuse albedo, related to the probability that light which enters the bulk will scatter back to the surface and exit, and $(\rho_l/\pi) \cos\theta_r$ is the standard Lambertian term for an ideal diffuse reflector. 
\begin{equation} W = \left(1-F(\theta_i,n_0/n)\right) \left( 1-F(\theta_o,n/n_0)\right)
\end{equation} 
is the Wolff factor \cite{Wolff:94}, which accounts for the Fresnel coefficients at the entrance and exit of the interface, and $\theta_o=\sin^{-1}(n_0/n \sin\theta_r)$ is the exit angle required for the light to be refracted into the final viewing angle $\theta_r$. The Oren-Nayar \cite{Oren:94} term $N\left(\theta_i,\theta_r,\phi_r;\gamma\right)$ accounts for the effects of surface roughness on the diffuse reflection, and involves an integral over both the microfacet distribution, $P$, as well as the shadowing and masking factor, $G$. This integral is carried out analytically for the case of the Trowbridge-Reitz distribution in \cite{Silva_thesis} and its form is used here:
\begin{equation}
N=G'(\theta_i)\,G'(\theta_r)\left( \mathcal{N}_0 - \tan\theta_i\tan\theta_r\cos\phi_r\,\mathcal{N}\right)
\end{equation}%
\begin{equation}
    \mathcal{N}_0\equiv\frac{1}{1-\gamma^2}-\frac{\gamma^2}{1-\gamma^2} \frac{\mathrm{arctanh}\left(\sqrt{1-\gamma^2}\right)}{\sqrt{1-\gamma^2}}
\end{equation}%
\begin{equation}
    \mathcal{N}\equiv\frac{\gamma^2}{2\,(1-\gamma^2)}-\frac{\gamma^2\,(2-\gamma^2)}{1-\gamma^2} \frac{\mathrm{arctanh}\left(\sqrt{1-\gamma^2}\right)}{\sqrt{1-\gamma^2}}
\end{equation}

A deficiency of this model is the prediction of no diffuse reflection when either of the global angles $\theta_i$ or $\theta_o$ is above the critical angle for total internal reflection, which becomes important when modelling reflectance into a medium of higher index than the PTFE, such as LXe. This prediction fails due to the fraction of microfacets with normal $\mathbf{\hat{n}'}$ such that $\theta'_i$ ($\theta'_o$) relative to the microfacet is below the critical angle, giving a non-zero value for the corresponding Wolff term. A full accounting of these effects can be achieved by including $W$ (as well as $P$ and $G$) into the integral performed to calculate $N$, but this becomes analytically intractable and cumbersome to evaluate numerically. The alternative used here is to use the analytic form of $N$ while calculating the incident (outgoing) part of $W$ as a numerical integral over $P$ only when the relevant global angle is beyond the critical angle. In practice, $F$ is very nearly a step function at the critical angle for the index ratios $n_0/n$ found for PTFE in LXe, therefore the relevant Wolff term is replaced by the fraction of microfacets from P that would result in $\theta'_i$ ($\theta'_o$) falling below the critical angle for total internal reflection. For typical parameters considered here, $1-F$ is within 5\% of unity when averaged over the relevant microfacet angle range, for all incident (outgoing) global angles. Therefore, we find this substitution to have a minimal effect on the final results.

The complete model is then given by: 
\begin{equation}\label{eqn:bridf}
\varrho\left(\theta_i,\theta_r,\phi_r;\rho_l,n_0/n,\gamma\right) = \varrho_D + \varrho_S,
\end{equation}
where $\varrho_D$ is defined by Eq.~\ref{eqn:rho_d} and $\varrho_S$ is defined by Eq.~\ref{eqn:rho_s} (or Eq.~\ref{eqn:rho_s_full} for samples exhibiting a specular spike). This model has the advantage of giving a complete prediction of the BRIDF at any incident or viewing angle, using only three to four parameters to describe the material in question: the diffuse albedo $\rho_l$, the index of refraction $n$, the surface roughness $\gamma$, and optionally the specular spike normalization factor $K$. Hence, measurements taken at a range of angles sufficient to constrain these three parameters should allow for extrapolation to the full hemisphere. 

Due to the finite beam size and PMT aperture size, the measured quantity in this work is not the BRIDF but rather its integral over a range of incident and viewing angles. To account for this, the model is averaged over a $4^\circ$ angular diameter circle in viewing angles $\theta_r$, $\phi_r$, to match the shape of the PMT aperture. A weighted average over incident angles $\theta_i$ following a Gaussian profile with standard deviation $\sigma = 2^\circ$ is also applied, to incorporate beam shape effects (Sec.~\ref{sec:setup}). Neither of these averaging methods significantly affects the shape of the reflectivity model unless a specular spike is observed, in which case its delta function shape is appropriately convolved with these averaging shapes. In such cases, the observed shape of the specular spike is well described by these nominal dimensions, without the need for a separate, independent fit parameter.
 
\section{Data Taking}\label{sec:data}

\subsection{Samples}\label{sec:samples}
To judge the effects of material type and surface preparation on reflectivity, a total of seven PTFE samples were measured in both vacuum and LXe. An additional Spectralon diffuse reflectance standard from Labsphere Incorporated was used for calibration (vacuum only). All samples were from molded PTFE powder. Some powder materials were chosen for direct comparison with the results of total reflectance measurements in LXe from \cite{Neves:2016}, including 807NX and NXT85 from Applied Plastics Technology and a sample of the material used in the LUX detector \cite{LUX-NIM}, molded 8764 PTFE from Technetics of an unknown powder material. These materials were found to have an estimated total reflectivity of >95\% \cite{Neves:2016,LUX-comprehensive}. Other samples were made from M17 and M18 PTFE powder from Daikin. M17 is the material used as a reflector in the LZ dark matter detector for both the outer skin region and the inner field cage \cite{LZ-TDR}.

All but three samples were prepared by machining the raw PTFE material into 2.54~cm diameter disks, with the surface to be measured carefully finished on a lathe for smoothness. This was found to provide a more uniform surface than milling. The sample of LUX material was taken directly from the LUX time projection chamber without further machining of the surface to be measured. One of the remaining samples was made from M17 cut to a thickness of 1~mm by the manufacturer by skiving. This sample is thus identical in both material and surface cut to the PTFE used to tile the inside of the LZ skin region \cite{LZ-TDR}. The final sample, made from M18, was machined on a lathe but subsequently polished with polishing paper of successively smaller grit sizes, down to 1~$\mu$m, until achieving a mirror polish by eye. This variety of surface finish methods allows for a study of the effect of surface roughness over a broad range, while including two samples (LUX and M17 skived) with surfaces that match those used in real LXe detectors ensures applicability to the domain of most interest. 

These samples were approximately 5~mm thick, with the exception of the skived M17 sample (1~mm). This is expected to be sufficiently thick for the reflectivity to be independent of thickness \cite{Haefner:2016}. Prior to measurement in vacuum, all samples were cleaned in an ultrasonic bath of ethanol to remove surface contaminants. Further handling was done using gloved hands, to maintain cleanliness. 

Samples were marked on the back side to ensure the same orientation in the sample rack for repeated measurements. Care was taken to ensure the location of the beam on the sample was the same for each measurement to within 2.5~mm, to minimize any systematic effects of reflectivity variation with position on the sample. 

\subsection{Data Taking Conditions}\label{sec:conditions}
For each sample, measurements were taken across a range of $\theta_i$ and $\theta_r$, chosen to fully constrain the model parameters. To avoid the effect of shadowing of the reflected light by the edge of the sample rack, the maximum $\theta_r$ for measurement is 85$^\circ$. Unless otherwise noted, the monochromator was set to 178~nm for data taking.

Measurements in LXe were taken over three runs of Xe fill and recovery, which primarily differed by the set of samples in the sample rack and the LXe purity (as judged by absorption at UV wavelengths). The same range of $\theta_i$ and $\theta_r$ was used as in vacuum for direct comparison. Many datasets show a sharply decreasing BRIDF above $\theta_r=80^\circ$; this is believed to be related to shadowing from the edge of the sample rack, a similar effect to that in vacuum, but exaggerated in LXe due to a small lensing effect of the cell (Sec.~\ref{sec:systematics}). For this reason, fits to the data are restricted to $\theta_r<80^\circ$ in LXe. 

Initial operating conditions were chosen to prevent bubbling in the LXe cell (as judged by observation through a side viewport) through rapid Xe circulation. The corresponding heat load caused a steady increase in Xe temperature and pressure. It was later discovered that Xe pressure has a much stronger effect on the reflectivity measurements than bubbling (Sec.~\ref{sec:systematics}). Because of this, most measurements are taken at a stable Xe pressure with some bubbling, though some early measurements instead prevent bubbling but vary in pressure from one dataset to the next. 

\subsection{Fitting}
Data at low incident angle primarily constrain the diffuse component and hence $\rho_l$ while the specular peak prominent at larger incident angles constrains $\gamma$ (primarily through its width) and $n_\mathrm{PTFE}$ through the peak's magnitude, position, and shape. Fit parameters are found by minimizing the $\chi^2$ of the model (Sec.~\ref{sec:model}) with respect to the combined dataset at all $\theta_i$ and $\theta_r$ for any given sample and measurement time. This is done with an adaptive gradient descent algorithm, which was found to outperform a simple grid search in both speed and fit quality. Fitting in LXe assumes an index of refraction of $n_\mathrm{LXe}=1.69$ \cite{Solovov:2003}; only the ratio $n_\mathrm{PTFE}/n_\mathrm{LXe}$ is observable in IBEX, hence deviations from $n_\mathrm{LXe}=1.69$, as for example when changing the LXe pressure (Sec.~\ref{sec:lxe}) are absorbed in the estimate of $n_\mathrm{PTFE}$.

\section{Results}\label{sec:results}
\subsection{Vacuum Measurements}\label{sec:vacuum}

\begin{figure}

  \centering\includegraphics[width=0.5\textwidth]{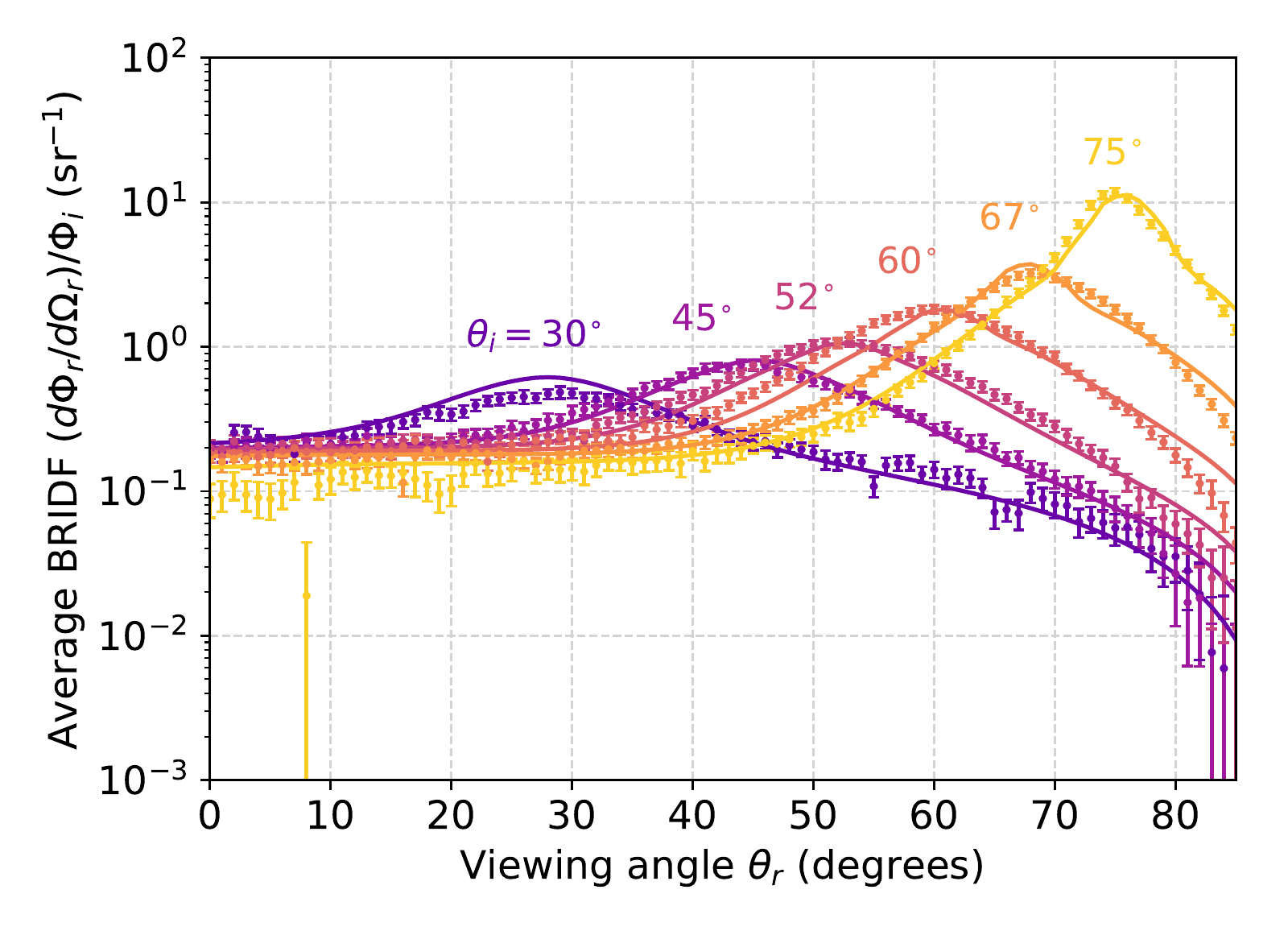}

\caption{Fitted BRIDF data for M17 skived sample in vacuum at 178 nm. Error bars are dominated by systematic errors (Sec.~\ref{sec:systematics}) and should not be considered independent. There is a $\pm2^\circ$ incident angle uncertainty that can shift each curve independently along the horizontal axis.}
\label{fig:vacuum-fit} 
\end{figure}

An example fitted dataset in vacuum is shown in Fig.~\ref{fig:vacuum-fit}. The error bars shown are a quadrature sum of statistical error (see Sec.~\ref{sec:setup}), constant estimated background subtraction uncertainty, and proportional error from beam intensity uncertainties. Because systematic effects dominate this error, the errors from one data point to the next should not be considered independent. A more detailed quantification of the systematic errors and their effects on extracted parameters is given in Sec.~\ref{sec:systematics}.

The primary features of the data shown in Fig.~\ref{fig:vacuum-fit} are representative of those seen on all samples. At low $\theta_r$, the diffuse component dominates, and decreases slightly with $\theta_i$ as more of the incident light is reflected at the surface rather than entering the bulk where it can scatter. A clear peak is seen at the specular angle for each $\theta_i$, with a magnitude that steadily increases with $\theta_i$ due to the higher Fresnel factor. Some of the smoother samples also show signs of a sharper peak at the highest $\theta_i$ values, due to the presence of coherent reflection (Sec.~\ref{sec:model}). 

\begin{table*}
\caption{Fit results in vacuum at 178~nm. $\rho_l$ is the diffuse albedo, $n_\mathrm{PTFE}$ is the PTFE refractive index, $\gamma$ is the surface roughness, and $K$ governs the specular spike normalization (Sec.~\ref{sec:model}). All reflective surfaces were finished on a lathe except for the polished and skived samples, and the sample taken directly from a LUX reflector.}
\label{tab:vac-fits}
\begin{tabular*}{\textwidth}{@{\extracolsep{\fill}}lrrrl@{}}
\hline
Sample & \multicolumn{1}{c}{$\rho_l$} & \multicolumn{1}{c}{$n_\mathrm{PTFE}$} & \multicolumn{1}{c}{$\gamma$} & \multicolumn{1}{c}{$K$} \\ 
\hline
LUX 8764        &  0.64 $\pm$ 0.03 & 1.39 $\pm$ 0.07 & 0.127 $\pm$ 0.011 & $\infty$ \\ 
M17             &  0.78 $\pm$ 0.10 & 1.59 $\pm$ 0.06 & 0.157 $\pm$ 0.010 & 10.9 $\pm$ 0.1 \\ 
M18              &  0.89 $\pm$ 0.05 & 1.59 $\pm$ 0.02 & 0.167 $\pm$ 0.002 & 13.9 $\pm$ 0.4 \\ 
NXT85           &  0.89 $\pm$ 0.07 & 1.59 $\pm$ 0.04 & 0.116 $\pm$ 0.007 & 11.2 $\pm$ 0.9 \\ 
807NX           &  0.91 $\pm$ 0.05 & 1.66 $\pm$ 0.05 & 0.148 $\pm$ 0.002 & $\infty$ \\ 
M18 polished    &  0.91 $\pm$ 0.05 & 1.54 $\pm$ 0.10 & 0.076 $\pm$ 0.012 & 3.6 $\pm$ 1.0 \\ 
M17 skived &  0.73 $\pm$ 0.10 & 1.70 $\pm$ 0.04 & 0.118 $\pm$ 0.004 & 8.0 $\pm$ 1.2 \\ 
\hline
\end{tabular*}
\end{table*}

A summary of the model fit parameters for vacuum measurements at 178~nm is given in Tab.~\ref{tab:vac-fits}. These results can be compared to fits using the same model from \cite{Silva:2009,Silva:2010,Silva_thesis}. A notable difference in the fits here is the consistently larger $\gamma$ values: 0.076-0.167 here as compared to a range of 0.049-0.066 in \cite{Silva:2010}, indicating rougher surfaces in this work. Albedos here tend to be slightly larger, though there is a wide range of values: 0.64-0.91 here versus 0.63-0.74 in \cite{Silva:2010}. Across samples, there is a mild anticorrelation between $\rho_l$ and $\gamma$, matching a similar (if somewhat stronger) finding in \cite{Silva:2009}. This suggests that smoother surfaces have improved diffuse reflectivity. Due to a partial degeneracy in the fits between $n_\mathrm{PTFE}$ and $\gamma$ for the vacuum measurements (70-80\% correlation) and an overall weaker dependence of the Fresnel factor on $n_\mathrm{PTFE}$ in vacuum, it is more difficult to gauge the PTFE index than in LXe. However, the range of values observed in the fits is comparable to those observed in prior work.

\subsection{LXe Measurements at Constant Pressure}\label{sec:lxe}

\begin{figure}

  \centering\includegraphics[width=0.5\textwidth]{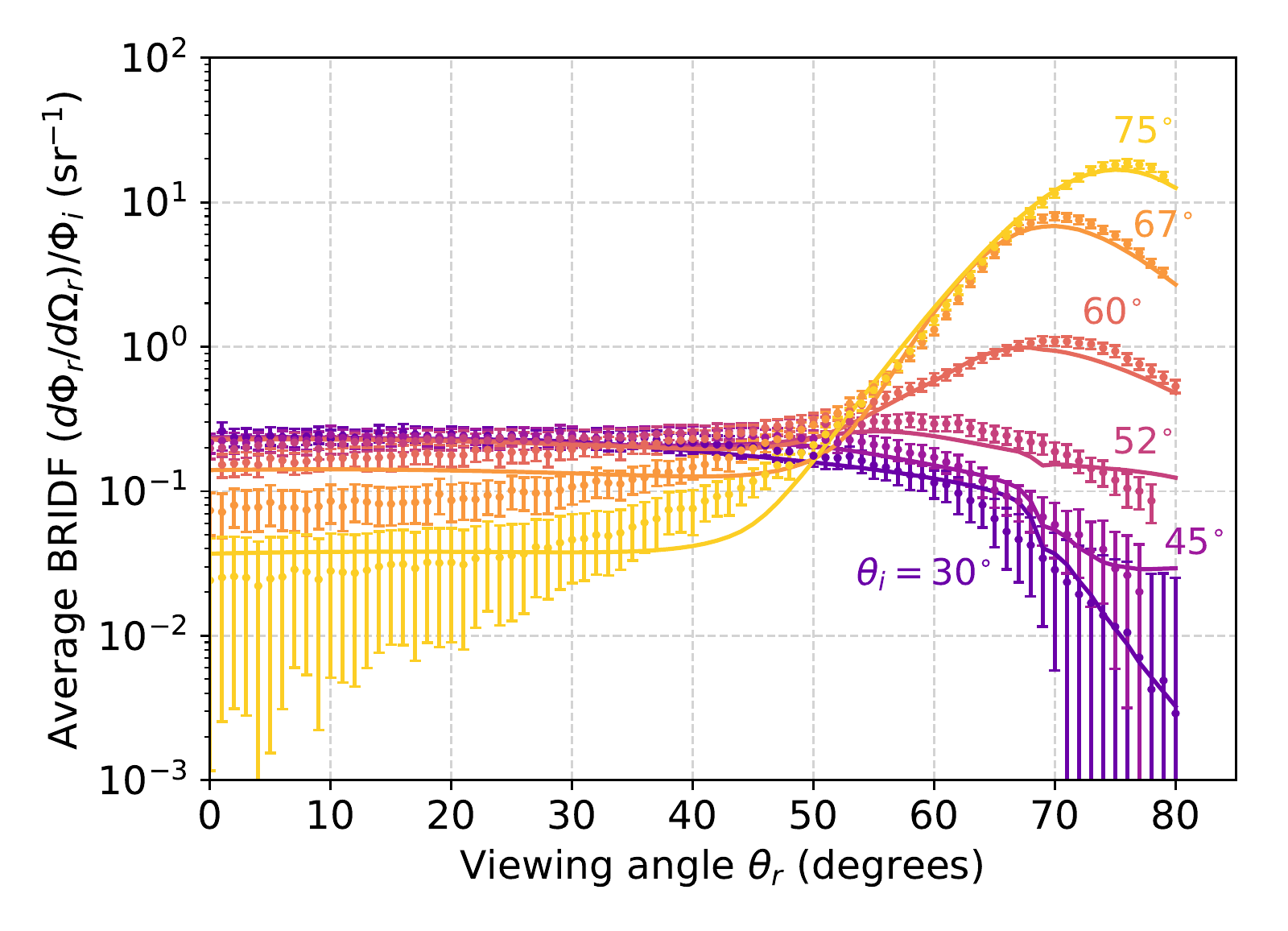}

\caption{Fitted BRIDF data for M17 skived sample in 0.25~barg LXe at 178~nm. Error bars are dominated by systematic errors (Sec.~\ref{sec:systematics}) and should not be considered independent. There is a $\pm2^\circ$ incident angle uncertainty that can shift each curve independently along the horizontal axis. A Gaussian distribution of refractive indices is used for a better fit to the data.}
\label{fig:lxe-fit-gauss-n} 
\end{figure}

\begin{figure}
  \centering\includegraphics[width=0.49\textwidth]{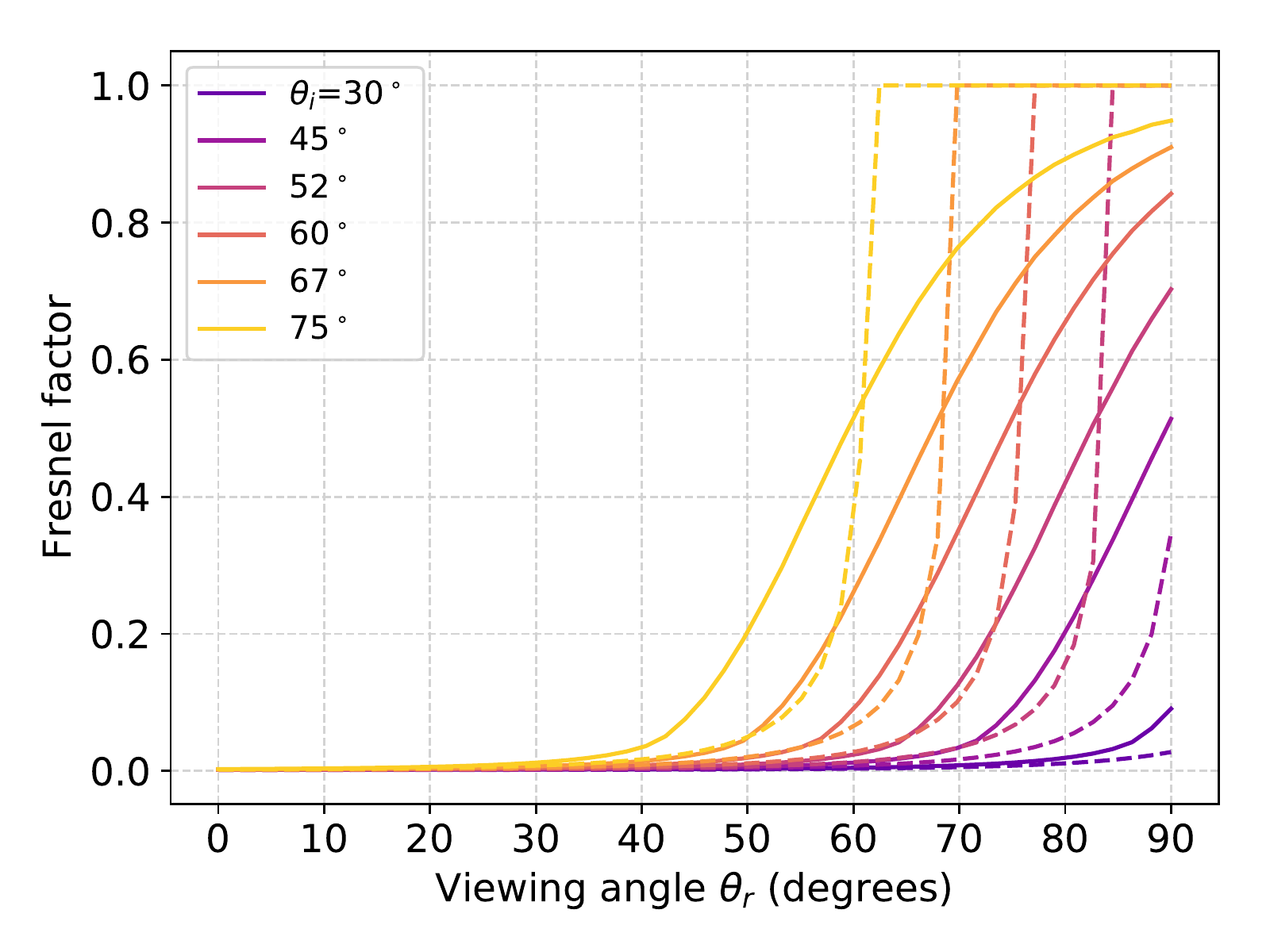}

\caption{Fresnel factor for specular reflection relative to the local microfacet versus viewing angle. Dashed curves use a single refractive index ratio ($n_\mathrm{PTFE}=1.57$, $n_\mathrm{LXe}=1.69$), resulting in a sharp turn-on near the critical angle for total internal reflection. Solid curves use a Gaussian range of indices, with Gaussian width $\sigma_n=0.07$. This leads to a more gradual transition near the critical angle. }
\label{fig:fresnel} 
\end{figure}

An example fitted dataset for PTFE immersed in LXe is shown in Fig.~\ref{fig:lxe-fit-gauss-n}. The larger error bars here compared to in vacuum come from a larger background subtraction uncertainty due to stronger reflections at the entrance to the cell (Sec.~\ref{sec:background}). These measurements in LXe share several key features across all samples. First, there are several signs that indicate $n_\mathrm{PTFE} < n_\mathrm{LXe}$, in which case the Fresnel factor, $F$, behaves similarly to a step function at the critical angle for total internal reflection (Fig.~\ref{fig:fresnel}). There is no clear specular peak below $\theta_i\sim60^\circ$, while at higher $\theta_i$ there is a dramatic rise in the specular component with a corresponding drop in the diffuse component. The specular peak is also shifted above the nominal specular angle ($\theta_r=\theta_i$) for curves at $\theta_i=60^\circ$ and $67^\circ$, which are near the critical angle. This peak shift phenomenon was also observed in \cite{Levy_thesis,Kaminsky_thesis}, with no explanation offered. It can be understood as arising from the sharp rising edge of $F$ for reflection off the local microfacets at a larger angle than the nominal specular angle, suppressing the specular peak until higher viewing angles. 

A second, unexpected feature of the data is a relatively long tail for the specular peak for $\theta_r<\theta_{peak}$; this cannot be adequately captured by alternative forms of the microfacet distribution function $P$, as they cannot undo the sharply-rising edge of $F$. Instead, a modification to $F$ that smooths out this edge is required.

Several such modifications were considered, including using a uniform range of indices $n_\mathrm{PTFE}$ as well as positing empirical sigmoid functions for $F$. The model which showed the best agreement with data was a Gaussian distribution of indices, with the width $\sigma_n$ left as an additional free parameter in the fits. This is the model used in all fits to the data in LXe. A comparison of the local Fresnel factor with and without this range of indices is given in Fig.~\ref{fig:fresnel}. Possible physical explanations for this, such as the different phases of PTFE giving rise to a range of refractive indices, are explored in Sec.~\ref{sec:wavelengths}.

\begin{table*}
\caption{Fit results in LXe at 178~nm. LXe pressures at the sample height during data taking were 0.83~barg for LUX~8764, 0.84-0.90~barg for M18, 0.44~barg for M18 polished, and 0.25~barg for all other samples. Fits assume $n_\mathrm{LXe}=1.69$. All reflective surfaces were finished on a lathe except for the polished and skived samples, and the sample taken directly from a LUX reflector.
*Only the M18 polished sample requires a specular spike term, with parameter $K=5.0 \pm 1.4$. 
}

\label{tab:lxe-fits}
\begin{tabular*}{\textwidth}{@{\extracolsep{\fill}}lrrrrl@{}}
\hline
Sample & \multicolumn{1}{c}{$\rho_l$} & \multicolumn{1}{c}{$n_\mathrm{PTFE}$} & \multicolumn{1}{c}{$\sigma_n$} & \multicolumn{1}{c}{$\gamma$} \\
\hline
LUX 8764        &  0.63 $\pm$ 0.06 & 1.58 $\pm$ 0.05 & 0.08 $\pm$ 0.05 & 0.15 $\pm$ 0.03 \\ 
M17             &  0.72 $\pm$ 0.09 & 1.55 $\pm$ 0.02 & 0.08 $\pm$ 0.03 & 0.21 $\pm$ 0.03 \\ 
M18             &  0.77 $\pm$ 0.05 & 1.57 $\pm$ 0.03 & 0.07 $\pm$ 0.01 & 0.17 $\pm$ 0.04 \\ 
NXT85           &  0.77 $\pm$ 0.05 & 1.575 $\pm$ 0.009 & 0.078 $\pm$ 0.003 & 0.110 $\pm$ 0.012 \\ 
807NX           &  0.74 $\pm$ 0.06 & 1.57 $\pm$ 0.03 & 0.10 $\pm$ 0.02 & 0.18 $\pm$ 0.03 \\
M18 polished*    &  0.91 $\pm$ 0.05 & 1.58 $\pm$ 0.02 & 0.063 $\pm$ 0.007 & 0.081 $\pm$ 0.006 \\ 
M17 skived &  0.76 $\pm$ 0.05 & 1.58 $\pm$ 0.02 & 0.063 $\pm$ 0.004 & 0.110 $\pm$ 0.014 \\ 
\hline
\end{tabular*}
\end{table*}

A summary of the model fit parameters for LXe measurements at 178~nm is given in Tab.~\ref{tab:lxe-fits}. With the exception of the M18 polished sample, including the specular spike of Eq.~\ref{eqn:rho_s_full} had a minimal effect on preliminary fits, hence the simpler Eq.~\ref{eqn:rho_s} is used. Albedos in LXe do not show a consistent increase over the corresponding values in vacuum when considering all samples, though the smoothest samples (M17 skived and M18 polished) do show evidence of such an increase. The mild anticorrelation between $\gamma$ and $\rho_l$ seen in vacuum measurements is also evident in LXe, most notably when comparing the M18 polished and unpolished samples. Due to the clear turn-on of the specular peaks near $\theta_i=60^\circ$, $n_\mathrm{PTFE}$ can be constrained quite tightly, and appears consistent across samples, with vacuum measurements in rough agreement. $\gamma$ is also consistent between vacuum and LXe measurements on the same sample. 

\subsection{LXe Pressure Dependence}

The effects of LXe pressure were also studied by taking reflectivity measurements at a range of stable pressures. These measurements showed a consistent suppression of the specular peak with increasing pressure at incidence angles of $\theta_i\geq52^\circ$, with the strongest effect near $60^\circ$. Initial tests included power measurements at each new stable pressure, which required resetting the sample angle using the rotation feedthrough each time. This introduces a potential systematic from the repeatability of setting the sample angle, to which the data are particularly sensitive near the critical angle for total internal reflection. 

\begin{figure}

  \centering\includegraphics[width=0.5\textwidth]{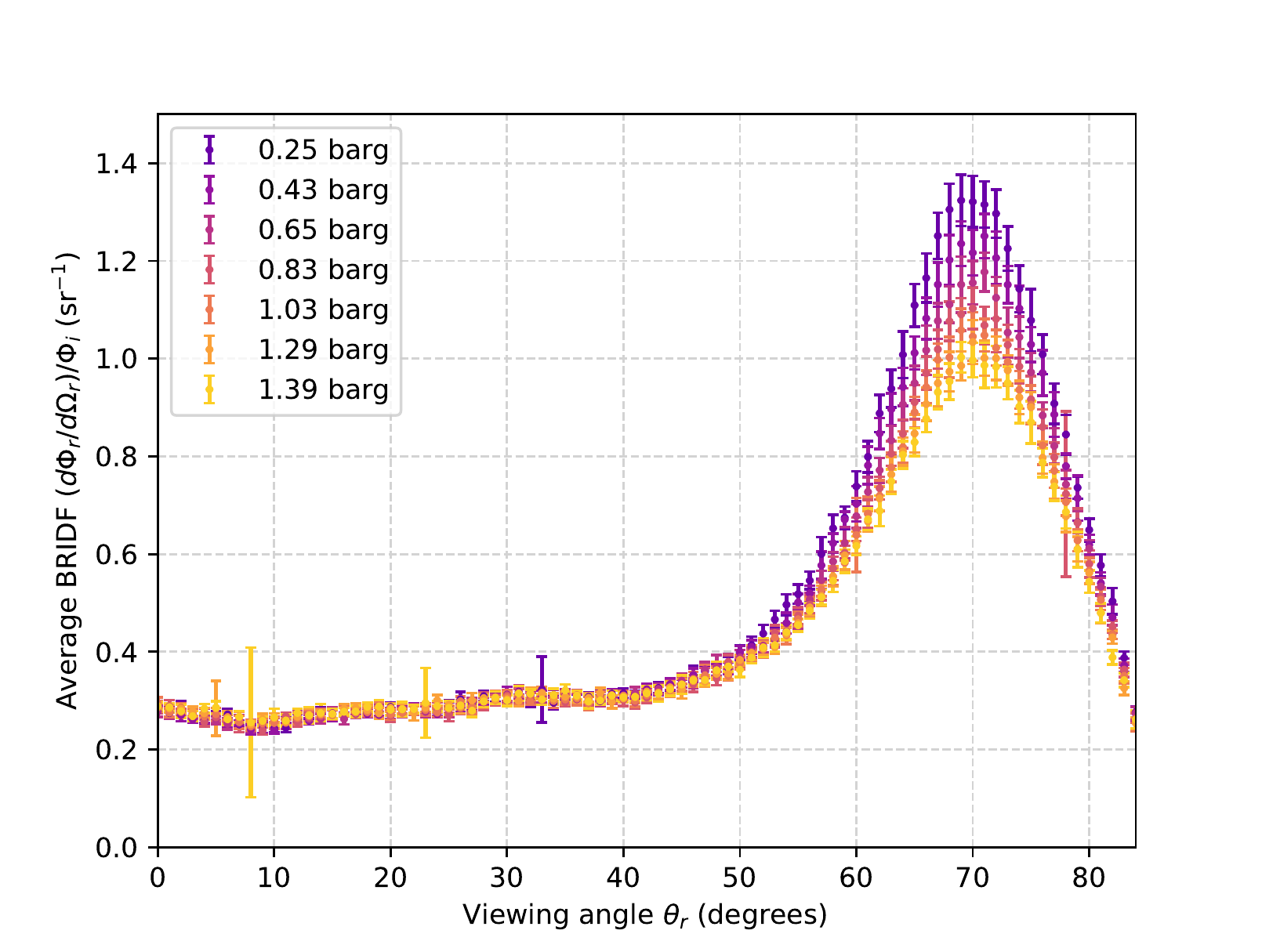}

\caption{Dependence of specular peak at $\theta_i=60^\circ$ on LXe pressure at the sample height for the M17 skived sample at 178~nm. Sample angle is fixed throughout, to avoid systematic uncertainties from the procedure to set $\theta_i$. The observed reduction in the specular peak with increased pressure is believed to be caused by the slightly lower LXe refractive index, which shifts the critical angle for total internal reflection up. Error bars are statistical only.}
\label{fig:pressure} 
\end{figure}

To avoid this, further measurements were taken with the rotation feedthrough fixed at all pressures, with the power measurement done before and after to bound the level of variation in the incident beam. One such test is shown in Fig.~\ref{fig:pressure}. The specular peak for the M17 skived sample at $\theta_i=60^\circ$ shows a decrease of $\sim$25\% at the highest pressure achieved of 1.39~barg relative to that at 0.25~barg. The pressures quoted here are inferred from the measured gas pressure plus the 0.05~bar additional pressure due to the column of LXe above the sample. Decreases of $\sim$25\% were also observed for the 807NX and M17 turned samples for $\theta_i=60^\circ$. 

This effect can be fully explained by the change in index of refraction of the LXe as its density varies with pressure. Prior measurements have been made of the LXe index at a range of optical wavelengths both at the triple point of -0.18~barg and at 178~K, corresponding to a pressure of 1.01~barg \cite{Sinnock:1969}. These data can be extrapolated to 178~nm using the Sellmeier equation as in \cite{Grace:2017}, with different coefficients at the two measured pressures. This corresponds to a decrease in index of $\sim$0.015 at the higher of these two pressures. To extend this to the pressures used here of 0.25-1.39~barg, one can, at first order, scale this index difference by the relative change in LXe density at these pressures, to get 0.019. 

Fitting the data in Fig.~\ref{fig:pressure} at 0.25~barg and then adjusting the LXe index by 0.019 while keeping all other model parameters fixed reproduces the decrease in the specular peak while leaving the shape relatively unchanged. The cause of this decrease in the model can be attributed to the shift in critical angle for total internal reflection due to a change in LXe index, which in turn causes a decrease in the Fresnel factor (see Fig.~\ref{fig:fresnel}). The magnitude of the effect estimated from this procedure (44\% of the 0.25~barg peak height) is somewhat larger than that seen in the data. However, a similar analysis can be done using data at a wider range of incident angles for both low and high pressure, to better constrain the model parameters than can be done using a single angle. These fits show a difference in $n_\mathrm{LXe}$ of 0.018, in good agreement with the change in index estimated above. All other fitted parameters are equal within errors between the datasets at different pressures. The model with this change in $n_\mathrm{LXe}$ also matches the smaller drop in the specular peak at $67^\circ$ and $75^\circ$ seen in the data.

Two additional checks were performed to support this interpretation. First, reflectivity measurements with fixed sample angle were repeated for both increasing and decreasing pressure, to check for effects from an unrelated time dependence. In both cases, the specular peak was larger for lower pressures, consistent with the changing index interpretation. Lastly, as a final check against the possibility of alignment changes as a result of pressure adjustment, the same test was performed using a mirror at $\theta_i=60^\circ$ in place of a PTFE sample, which provides the most sensitive check of the alignment as it preserves the narrow incident beam shape. This showed changes in the reflected peak height of <5\%, well below the size of the pressure effect at $60^\circ$.

Despite the high significance of this effect for measurements within the plane of incidence, the hemispherical reflectance (Sec.~\ref{sec:refl}) is not expected to show a strong pressure dependence. This is because the effect is small for incident angles far from the critical angle, and the specular component accounts for less than half of the hemispherical reflectance at those angles. Using fits at two different pressures for the M17 skived sample and the extrapolation procedure described in Sec.~\ref{sec:refl}, the maximum difference in hemispherical reflectance at any incident angle is <10\% -- within the uncertainty on the extrapolated values. In a real LXe detector, where light is collected across a broad range of incident angles, this difference is expected to be smaller still.

\subsection{Measurements at Other Wavelengths}\label{sec:wavelengths}
\begin{figure*}

  \centering\includegraphics[width=0.49\textwidth]{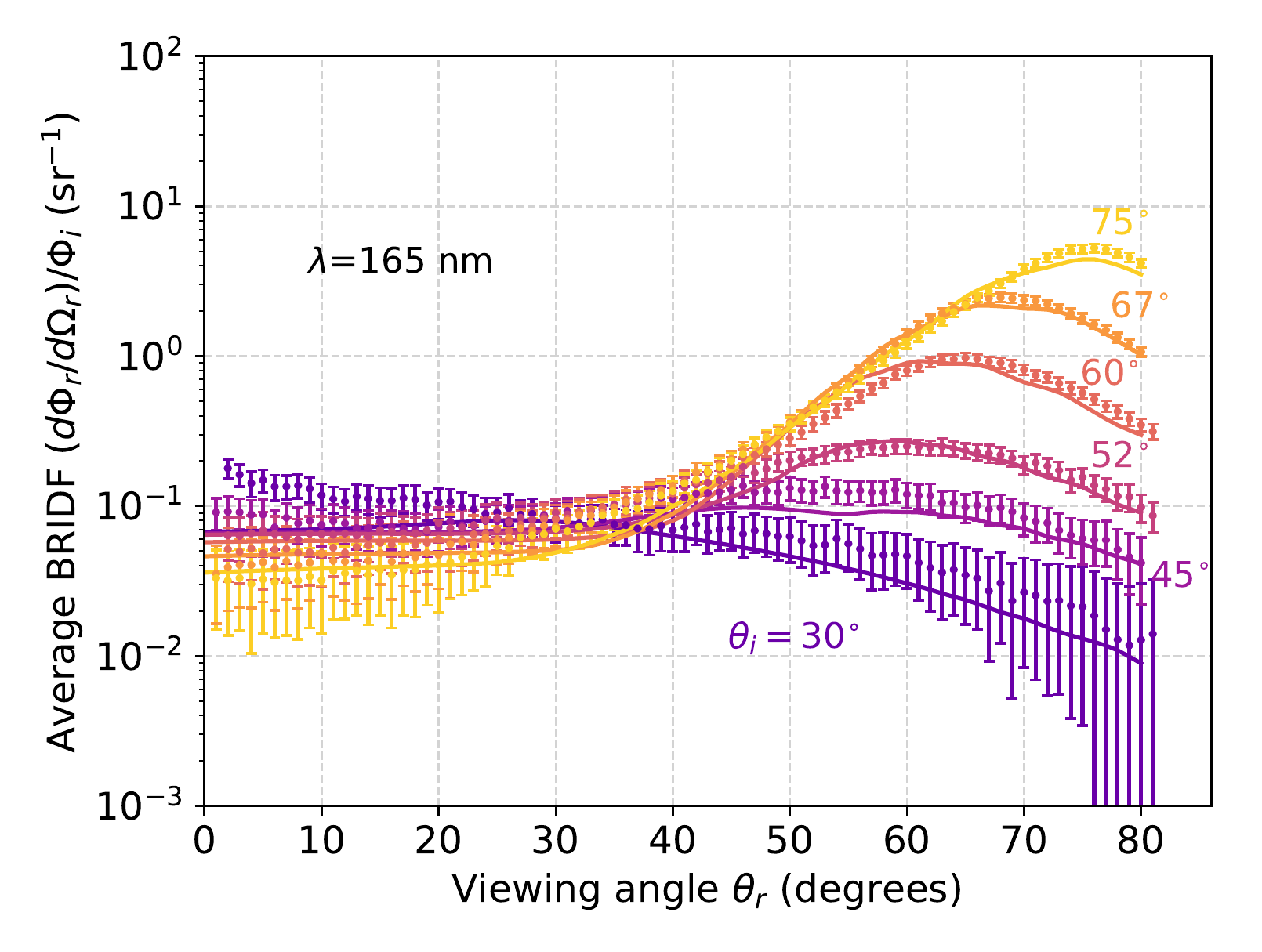}
  \centering\includegraphics[width=0.49\textwidth]{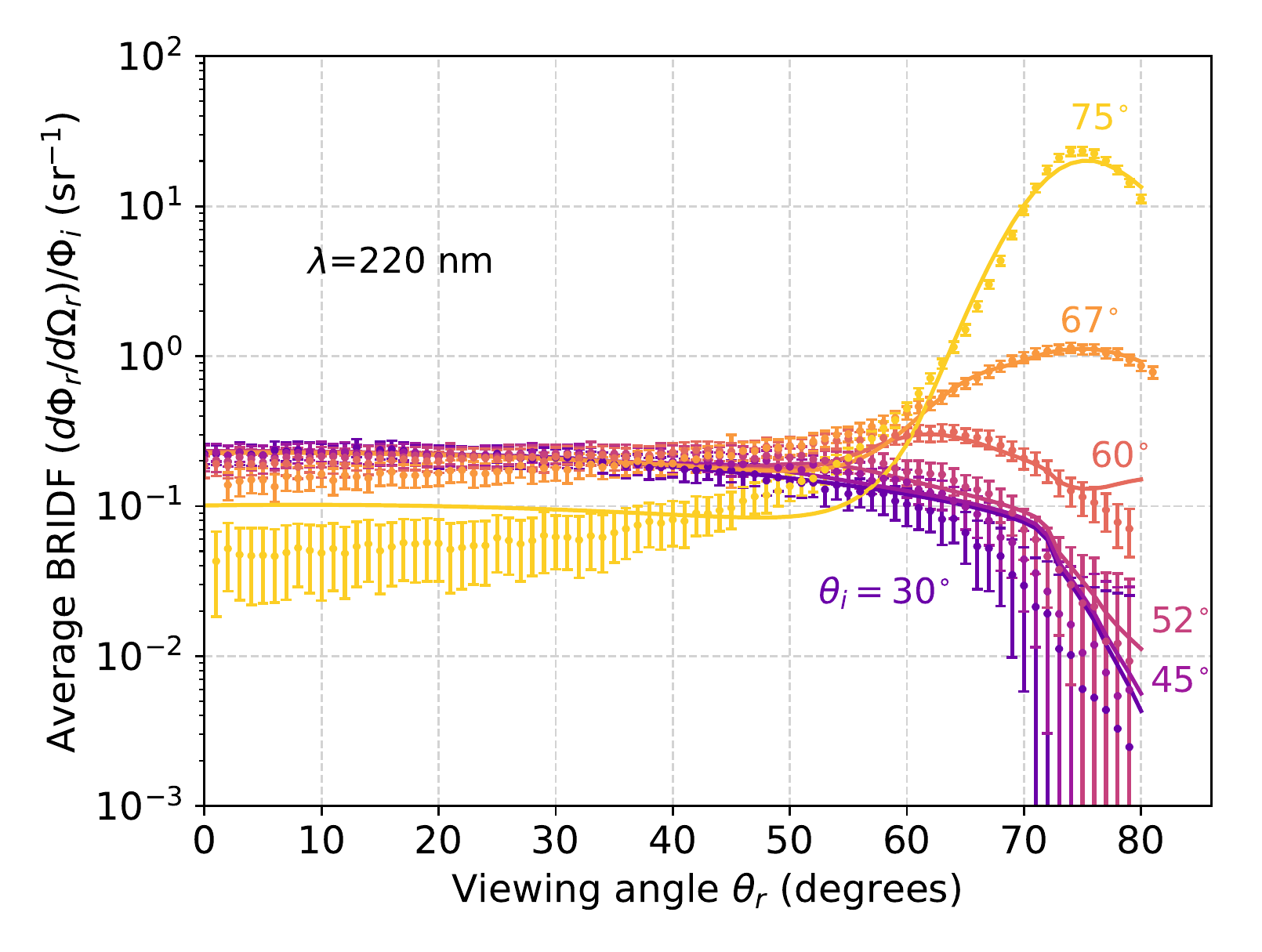}
  \centering\includegraphics[width=0.49\textwidth]{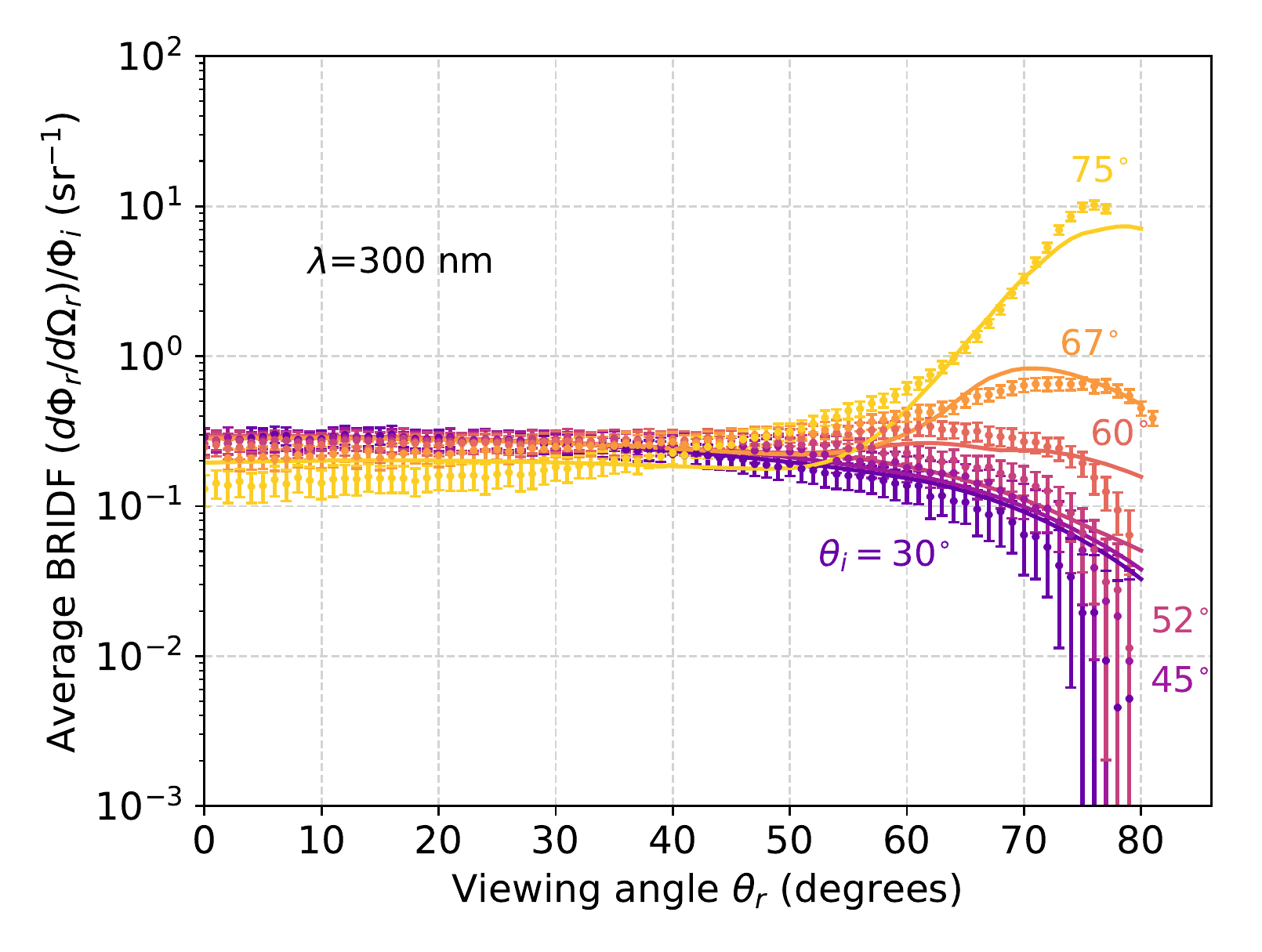}
  \centering\includegraphics[width=0.49\textwidth]{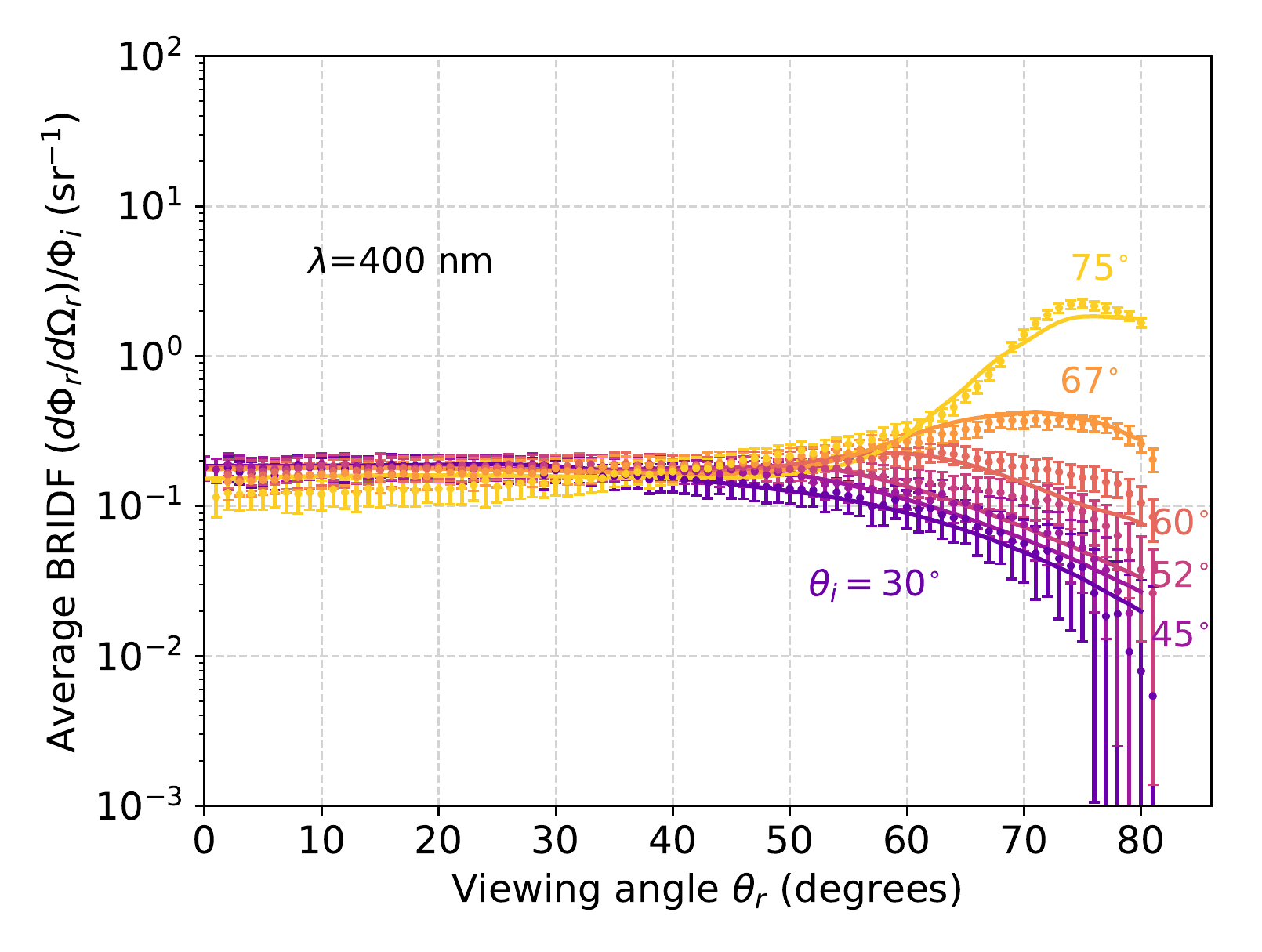}

\caption{Fitted BRIDF data for the M17 skived sample immersed in LXe at other wavelengths. Error bars are dominated by systematic errors (Sec.~\ref{sec:systematics}) and should not be considered independent. There is a $\pm2^\circ$ incident angle uncertainty that can shift each curve independently along the horizontal axis.}
\label{fig:wavelength} 
\end{figure*}

Further measurements of reflectivity at wavelengths other than 178~nm were performed using the M17 skived sample immersed in LXe (Fig.~\ref{fig:wavelength}). Fit parameters are presented in Tab.~\ref{tab:wavelength-fits}. These indicate a substantial drop in diffuse reflectance at 165~nm, likely due to the presence of a PTFE absorption peak near 161~nm \cite{Seki:1990}. At longer wavelengths, the diffuse reflectance is relatively flat, with a slight peak near 300~nm. This peak was also observed in the vacuum measurements of the same sample. It was further duplicated in calibrated total reflectance measurements of this sample performed by Labsphere Incorporated in vacuum at wavelengths above 250~nm using an integrating sphere (more details in Sec.~\ref{sec:refl}). Prior measurements of PTFE reflectivity in vacuum \cite{Silva:2009} did not indicate such a feature. Therefore, it may be sample-specific, possibly related to fluorescence from trace impurities in the PTFE \cite{Steigman:1993}. The best fit albedo at 400~nm appears lower than at other wavelengths, but the error is substantially higher due to increased uncertainty from background subtraction, as there is significantly more stray light for visible wavelengths.

\begin{table*}
\caption{Fit results in LXe at a range of wavelengths on the M17 skived sample at 0.25~barg. Values for $n_\mathrm{LXe}$ are calculated using the Sellmeier equation as explained in the text and are not fit parameters.
}

\label{tab:wavelength-fits}
\begin{tabular*}{\textwidth}{@{\extracolsep{\fill}}lrrrrl@{}}
\hline
Wavelength & \multicolumn{1}{c}{$\rho_l$} & \multicolumn{1}{c}{$n_\mathrm{LXe}$} & \multicolumn{1}{c}{$n_\mathrm{PTFE}$} & \multicolumn{1}{c}{$\sigma_n$} & \multicolumn{1}{c}{$\gamma$} \\
\hline
165 nm        &  0.22 $\pm$ 0.07 & 1.901 & 1.92 $\pm$ 0.06 & 0.29 $\pm$ 0.06 & 0.15 $\pm$ 0.02 \\ 
178 nm        &  0.76 $\pm$ 0.05 & 1.690 & 1.58 $\pm$ 0.02 & 0.063 $\pm$ 0.004 & 0.11 $\pm$ 0.01 \\ 
220 nm        &  0.74 $\pm$ 0.05 & 1.504 & 1.444 $\pm$ 0.009 & 0.03 $\pm$ 0.05 & 0.083 $\pm$ 0.005 \\ 
300 nm        &  0.94 $\pm$ 0.09 & 1.430 & 1.44 $\pm$ 0.14 & 0.08 $\pm$ 0.05 & 0.11 $\pm$ 0.02 \\ 
400 nm        &  0.60 $\pm$ 0.18 & 1.404 & 1.54 $\pm$ 0.01 & 0.13 $\pm$ 0.05 & 0.138 $\pm$ 0.007 \\ 
\hline
\end{tabular*}
\end{table*}

The fitted PTFE refractive index drops with increasing wavelength from 165~nm to 300~nm, as expected due to the UV absorption peak at 161~nm. It then appears to increase above 300~nm, a feature also observed with low significance in vacuum measurements of the same sample as well as in prior work \cite{Silva:2009}.

A particularly striking difference for the 165~nm data is the broad nature of the specular peaks, captured in the model by a large range of indices, $\sigma_n$. A possible explanation for this is the finite range of wavelengths exiting the monochromator (Sec.~\ref{sec:setup}), with $\sigma_\lambda=1.5$~nm for the 165~nm measurements. However, this is insufficient to explain the range of indices seen if only considering the change in $n_\mathrm{LXe}$ with wavelength, using the Sellmeier equation and prior LXe index measurements as explained in Sec.~\ref{sec:lxe}. $n_\mathrm{PTFE}$ is expected to change even more strongly with wavelength in this region, due to the absorption peak near 161~nm. To account for this, the PTFE indices from Tab.~\ref{tab:wavelength-fits} were fit to the Sellmeier equation with a single absorption line at 161~nm: 
\begin{equation}\label{eqn:n-vs-lambda}
n_\mathrm{PTFE}(\lambda) = \sqrt{\left(a_0+a_{UV} \lambda^2\right)/\left(\lambda^2-\lambda_0^2\right)},
\end{equation}
where $a_0$ and $a_{UV}$ are fit parameters and $\lambda_0=161$~nm. 

Though Eq.~\ref{eqn:n-vs-lambda} does allow for a broader index range, it is still insufficient to fully explain the $\sigma_n$ values from the fits. This remains true for slight variants in the Sellmeier fit such as adjusting $\lambda_0$ and excluding the visible wavelength measurements of $n_\mathrm{PTFE}$. To match the value of $\sigma_n$ from fitting at 165~nm (178~nm), it is required to increase the wavelength range $\sigma_\lambda$ by a factor of $\sim$5 ($\sim$10), suggesting a different physical origin for this effect. A possible alternative explanation is PTFE's composition of both amorphous and crystalline phases, which have measurably different refractive indices in the visible wavelength range \cite{Fluoropolymers,Groh-1991}. The model in vacuum is relatively insensitive to adding a spread in indices $\sigma_n$, making the measurements here unable to distinguish this possibility.

\subsection{Extrapolated Reflectance}\label{sec:refl}

All reflectivity measurements were taken in the plane of incidence. However, the model presented in Sec.~\ref{sec:model} allows for extrapolation of these measurements to the full hemisphere. This model predicts a relatively steep drop-off in specular reflectivity away from the $\phi_r=0$ plane, which has been qualitatively observed in prior measurements in LXe \cite{Kaminsky_thesis}. The hemispherical reflectance of any given sample can thus be estimated as a function of the angle of incidence of incoming light through numerical integration of the BRIDF in Eq.~\ref{eqn:bridf}:
\begin{equation}\label{eqn:R-extrap}
    R\left(\theta_i\right) = \int \frac{1}{G}\varrho\left( \theta_i,\theta_r,\phi_r \right) \sin{\theta_r}d\theta_r d\phi_r.
\end{equation}
Here the shadowing and masking factor $G$ is removed from the BRIDF, to account for the fact that shadowed or masked light may ultimately be reflected.

\begin{figure}

  \centering\includegraphics[width=0.5\textwidth]{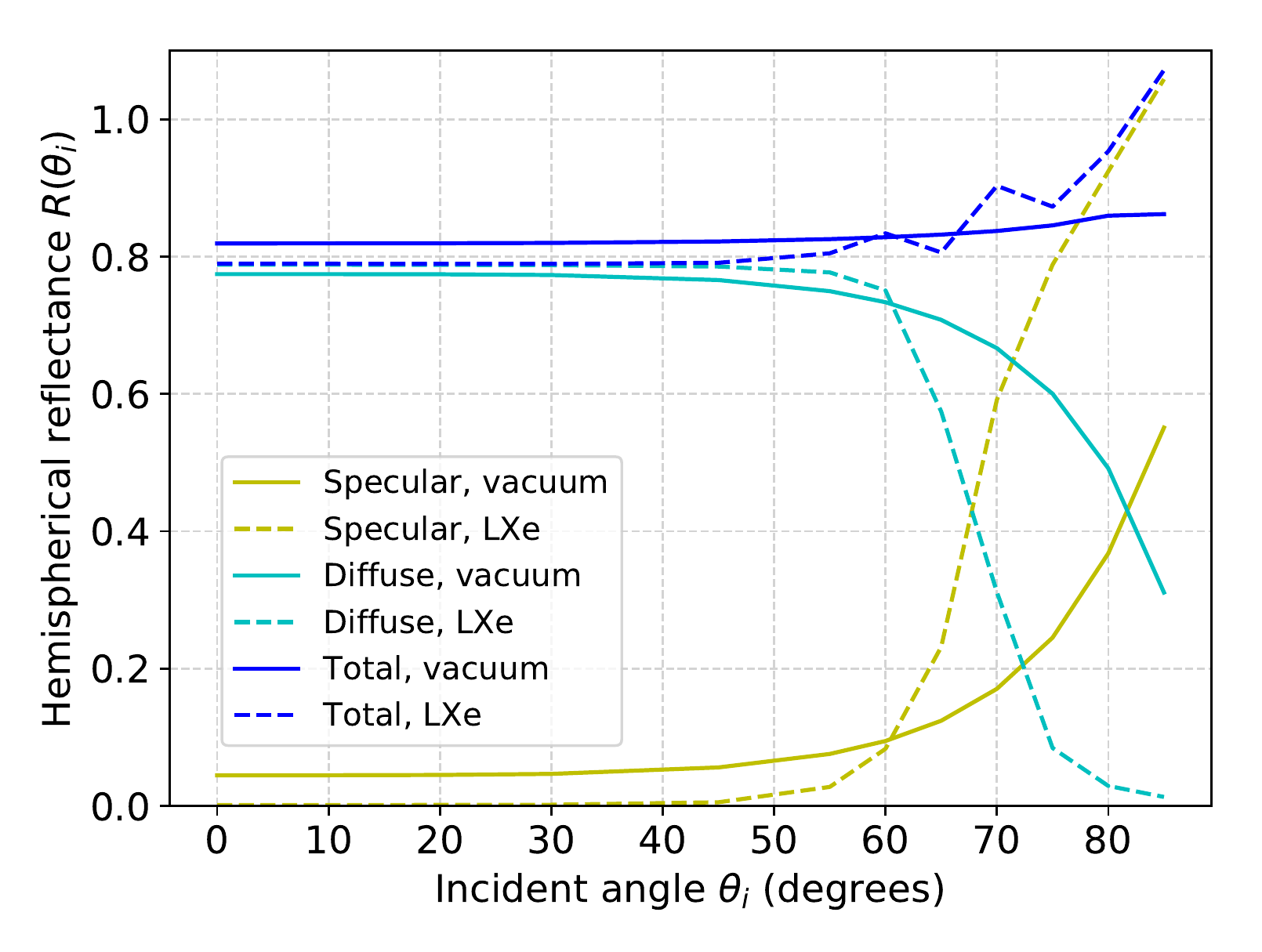}

\caption{Hemispherical reflectance versus incident angle for the M18 polished sample in vacuum and in 0.25~barg LXe at 178~nm. Extrapolation is done using the BRIDF model described in Sec.~\ref{sec:model} and best fit parameters for each dataset. The specular and diffuse components are separated for illustration. Uncertainties on the total reflectance range from 11\% at low angles up to 18\% at 85$^\circ$, with further details in the text.}
\label{fig:hemispherical-reflectance} 
\end{figure}

An example of $R\left(\theta_i\right)$, calculated for the M18 polished sample in vacuum and in LXe, is given in Fig.~\ref{fig:hemispherical-reflectance}. The reflectance versus incident angle remains fairly constant up to $\sim$60$^\circ$ in both vacuum and LXe, and is dominated by the diffuse component throughout that range. At higher angles in LXe, the specular component begins to dominate and the total reflectivity increases, nearing 100\%, as total internal reflection becomes relevant. This transition is smoothed out due to both the surface roughness and the range of refractive indices used. In vacuum, the specular component is less suppressed at low $\theta_i$ than in LXe due to the larger index mismatch. This sample is believed to best match the smooth surface of the molded PTFE used for the inner reflectors of a LXe TPC such as LZ \cite{LZ-TDR}, though other samples with rougher surfaces showed somewhat lower reflectances (Sec.~\ref{sec:lxe}).

Uncertainties on this total reflectance come from errors on the numerical integration (significant only near $\theta_i = 90^\circ$, causing the estimated total reflectance to exceed 1.0) and comparison with calibrated reflectance measurements performed by Labsphere Incorporated. For this comparison, three of the samples used here were sent to Labsphere for measurement in an integrating sphere at wavelengths from 250~nm up through the entire visible range. In this setup, the sample is put inside of a highly reflective sphere; light incident at 8$^\circ$ from the normal reflects off the sample, then undergoes a series of reflections off the enclosure walls until escaping through an exit aperture. The total light exiting the aperture is compared against similar measurements from a calibrated standard of known reflectance to gauge the sample's reflectance. The calibrated values were compared to that from Eq.~\ref{eqn:R-extrap} for $\theta_i=8^\circ$ using data in vacuum at select wavelengths in the same range. The extrapolated model showed a systematic bias toward lower values of 8\% on average, with an RMS difference between the two measures of 11\%. This 11\% provides a rough estimate of the uncertainty on the hemispherical reflectance, increasing to 18\% at the highest angles when combined with numerical integration errors. 

As an additional cross check, a Spectralon diffuse reflectance standard from Labsphere, calibrated to 97\% reflectance at 400~nm and $\theta_i=8^\circ$, was also measured in IBEX in vacuum at 400~nm and different stages of data taking. These measurements showed some evidence of degradation over time, which was confirmed by a subsequent recalibration by Labsphere, indicating a drop to $\sim$87\% at 400~nm. Their standard washing and sanding procedure, designed to restore the surface, improved this value to 96\%. The cause of this degradation is unknown, but may also lead to a reduced reflectance of other samples measured. The fact that the sample with the lowest reflectance (LUX~8764) was also handled the most and its surface prepared first further supports the possibility of reflectance degradation over time.

\section{Systematics}\label{sec:systematics}

Potential sources of systematic uncertainty on the reflectivity measurements were investigated, with their effect on the extracted fit parameters quantified where possible. For those systematics with an independent constraint, each fit parameter's uncertainty is given by its change between the best fit value and the $\pm1~\sigma$ variation on the systematic. 

One such systematic comes from uncertainty on the incident beam intensity, due to fluctuations over time ($\sim2$\%) and uncertainties related to the beam profile. As noted in Sec.~\ref{sec:power}, the full beam does not enter the PMT aperture during a power measurement, requiring a correction of the beam intensity according to the modeled beam profile. The uncertainty on this correction translates directly into a scaling of the reflected intensities.

Uncertainty in the background from scattered light also primarily affects $\rho_l$, as the count rates near the specular viewing angle are orders of magnitude larger than the background. This is most notable in LXe measurements with poor Xe purity which require high beam intensity incident at the entrance to the cell. The background subtraction procedure described in Sec.~\ref{sec:background} corrects for this, but an additional scaling factor in the background profile is introduced to account for changes between background measurements. This scaling factor is adjusted according to the maximum range of relative UV transmission measurements at 178~nm through the LXe during the data-taking run of interest. In extreme cases, this range can reach as low as 0.5 and as high as 2.5. Fits spanning the range of these scaling factors are then used to judge the variation on the fit parameters.

For systematics where an independent constraint was not available, parameter uncertainties are quantified by adjusting the model (or data, as appropriate) according to a change in the systematic, then dividing the change in each fit parameter by the square root of the change in the reduced $\chi^2$ to get the $1\sigma$ uncertainty: $\sigma_p = \Delta p/\sqrt{\Delta\chi^2/n}$ for any given parameter, $p$. This assumes that the $\chi^2$ profile for each fit parameter is quadratic in the relevant systematic. These systematics are then summed in quadrature, including the constrained systematics, to get the final parameter uncertainties.

Of these systematics, one of the most significant is the $\pm 2^\circ$ uncertainty on each of the incident angles, $\theta_i$ (Sec.~\ref{sec:mirror}). This is accounted for by comparing the combination of $\theta_i$ shifts that provides the best fit with the case of no shifts. The determination of correct incident angles in vacuum is typically straightforward due to the specular peak aligning with the mirror reflection angle. In LXe, specular peak shifts add complexity which translates primarily to increased uncertainty on $n_\mathrm{PTFE}$ from this systematic.

A further unconstrained systematic is present for the few datasets which showed evidence of the incident beam being masked by the sample rack (Sec.~\ref{sec:mirror}). In these cases, the $\chi^2$ comparison was done between the case with beam intensity scaled for high $\theta_i$ according to mirror measurements and without this scaling. For LXe measurements, an additional systematic compares fits with the Gaussian range of indices described in Sec.~\ref{sec:lxe} to that with a single index, primarily to quantify any possible bias on $n_\mathrm{PTFE}$ from this change in model as compared with vacuum. By necessity, there is no corresponding error on $\sigma_n$, as the single index model does not include this parameter. 

Additional systematics specific to operation in LXe were considered and found to have negligible effect on the data. Refraction at the entrance and exit of the cell, for rays not entering or exiting perfectly normal to the cell surface, was minimal and accounted for by the mirror angle calibration. Absorption in the LXe or the cell is accounted for in the power measurements, as the path length through each is the same whether passing directly through or reflecting off a sample. As confirmation of this, for the third run in LXe, during which the purity as judged by UV transmission worsened noticeably over time, measurements did not show any changes beyond what is quantified in the background subtraction systematic above. Scattering in the LXe is also accounted for, through background subtraction measurements. Lastly, the presence of bubbles in the LXe passing through the plane of incidence had no noticeable effect on reflectivity measurements, as judged by initial tests during operation with and without bubbles present.

A remaining systematic that could not be rigorously quantified was the possibility of changes in reflectivity over time. This was studied by measuring some samples during various stages of data collection in both LXe and vacuum, and across a range of LXe purities. In this limited study, the only pattern observed was mild evidence of an increase in diffuse reflectivity over exposure time in LXe. However, this is difficult to disentangle with changes in pressure and the effects of decreasing LXe purity over time, requiring increased incident light intensity and correspondingly higher background from stray reflections at the cell. This evidence appears at a similar size to the background subtraction systematic error, and hence further research is needed to determine whether this effect is real or due to correlated systematics. Because of this, parameter estimates in Tab.~\ref{tab:vac-fits}-\ref{tab:wavelength-fits} are from single measurements.

\section{Discussion and Conclusion}\label{sec:conclusion}

IBEX has measured the angular dependence of PTFE reflectivity in liquid xenon. These measurements indicate a strong dependence on incident angle, deviating from purely diffuse reflection. In particular, total internal reflection is observed for high angles of incidence, resulting in a specular component that dominates the reflectivity at these high angles and provides an enhancement in hemispherical reflectance when compared with equivalent measurements in vacuum. Near the critical incident angle for total internal reflection, the specular peak shows a high sensitivity to Xe pressure. This is readily explained by the change in Xe index of refraction with the density, and does not lead to a significant change in total reflectance for the majority of the range of possible incident angles. 

The onset of total internal reflection is observed to be a more gradual function of viewing angle than expected for a single ratio of refractive indices at the LXe-PTFE boundary. Modeling the index ratio as a Gaussian distribution rather than a single value gives a good fit to the data, and may be explained by multiple phases of the PTFE. Xe bubble formation at the Xe-PTFE boundary (in principle an explanation for the high observed PTFE reflectivity) is disfavored, as the large difference in index between Xe gas and PTFE or LXe would result in a secondary specular peak which is not observed.

There is some tension with prior measurements which indicate a higher total reflectance than observed here \cite{Neves:2016,LUX-comprehensive}. There are several possible sources of this discrepancy. First, the sample surface preparation is different, with generally rougher surfaces in this work. This is likely relevant due to the observed negative correlation between albedo and surface roughness. Surface degradation, seen here in a calibrated reference sample, may also play a role, and may result from trace contamination. A broader wavelength distribution for LXe scintillation light as compared to the light source used in IBEX is expected to shift the average refractive index somewhat and may also affect the albedo. Fluorescence of the PTFE in the visible, due to impurities, has been observed in prior work \cite{Gachkovskii:1969,Weidner:81,Steigman:1993} and may differ between this and other setups. In particular, an enclosed PTFE geometry such as those in \cite{Neves:2016,LUX-comprehensive} may enhance fluorescence effects \cite{Saunders:76,Shaw:08}. Angle-resolved measurements in \cite{Levy_thesis,Kaminsky_thesis} also indicate lower total reflectance than expected, suggesting the difference in measurement geometry may be relevant.

Lastly, direct comparison across geometries is difficult due to the observed dependence of reflectivity on incident angle, which creates a partial degeneracy between the angular distribution of incident light and the total surface reflectivity. For geometries supporting multiple reflections, this degeneracy extends to the reflected light distribution as well. The observed phenomenon of total internal reflection will concentrate light toward higher incident angles. It is therefore inadequate to assume a purely diffuse PTFE reflectivity model in Xe detectors. Future work incorporating a realistic distribution of incident angles along with BRIDF measurements presented here is required to accurately simulate light propagation in such detectors. Additional studies of PTFE fluorescence in this context, and of possible effects of extended LXe exposure on the PTFE BRIDF, will help further understanding of the light collection of Xe detectors.

\begin{acknowledgements}
The authors would like to thank Alexandre Lindote, Francisco Neves, and Cl\'audio Silva for helpful correspondence. The Ocean Optics spectrometer, McPherson monochromator, deuterium light source, lamp power supply, and monochromator motor controller are kindly on loan from Dr.~K.~Rielage and Los Alamos National Laboratory. We gratefully acknowledge the technical and engineering staff at LBNL for their support. This work was supported by the Laboratory Directed Research and Development Program of Lawrence Berkeley National Laboratory under U.S. Department of Energy Contract No. DE-AC02-05CH11231. 
\end{acknowledgements}


\appendix

\section{Summary of alignments and corrections}\label{app}
See Tab.~\ref{tab:alignments} and \ref{tab:calibrations}.

\begin{table*}
\caption{Summary of apparatus alignment procedures}

\label{tab:alignments}
\begin{tabular*}{\textwidth}{@{\extracolsep{\fill}}P{2cm}P{5cm}P{4cm}P{1.5cm}P{3.25cm}}
\hline
Procedure & Summary & Frequency & Conditions & Effect \\
\hline
Collimator positioning & Verify that beam through collimator strikes thin piece along sample rack axis & Once prior to all measurements & Atmosphere & Aligns collimated beam to axis of rotation of samples \\ 
\hline
Monochromator positioning &  Adjust orientation of monochromator towards maximum photon rate through collimator & Once prior to all measurements & Vacuum & Aligns monochromator to collimator for maximum incident power \\ 
\hline
Rotation stage centering about sample rack & Lower sample rack into rotation stage bore and check it is centered by eye. Verify correspondence of rotation of incident angle on sample transfer arm dial with shift in viewing angle of beam reflected by mirror & Rotation stage adjusted twice, before and after first LXe run; mirror measurements taken each time vacuum chamber conditions changed & All & Aligns PMT rotation axis with sample rotation axis \\ 
\hline
Fused silica cell centering about sample rack & Adjust orientation of bellows connected to cell until sample rack rotates freely in column; monitor fluctuations in beam position over course of experiment & Bellows adjusted once prior to all measurements. Beam position recorded in regular power measurements. & Adjusted at atmosphere, monitored in vacuum and LXe & Minimize systematics due to refraction at cell \\ 
\hline
Sample height setting & Hold sample rack at grazing angle to beam, adjust height towards maximum photon rate (beam incident on notch at sample equator); mark corresponding position on transfer arm. & Once prior to all measurements & Vacuum & Minimize systematics due to beam missing samples or being blocked by rack \\
\hline
PMT height setting & Reduce PMT aperture to minimum, scan about beam transmitted through collimator, adjust towards height with maximum photon rate & Twice; prior to all measurements and after replacement of damaged PMT base following second LXe run & Atmosphere & Set PMT to scan in plane of incidence \\ 
\hline
\end{tabular*}
\end{table*}

\begin{table*}
\caption{Summary of calibration and correction procedures}

\label{tab:calibrations}
\begin{tabular*}{\textwidth}{@{\extracolsep{\fill}}P{2cm}P{5cm}P{4cm}P{1.5cm}P{3.25cm}}
\hline
Procedure & Summary & Frequency & Conditions & Effect \\
\hline
Power measurement & Lift sample rack clear of beam, scan PMT about beam and record maximum photon rate & Whenever deuterium lamp is turned on, monochromator wavelength is changed, or any other change in conditions is expected & All & Determines uncorrected $\Phi_i$. Fluctuations of 2\% contribute to systematic uncertainty in $\Phi_i$\\ 
\hline
Beam modeling &  2D beam profile from measurements at atmosphere is fit with Gaussian; Monte Carlo convolution of this model with circular aperture is compared to power measurements in vacuum and LXe, with magnification factor applied in horizontal direction in LXe case due to refraction & Model from single 2D measurement validated against many power measurements & Atmosphere & Determines correction factor (in vacuum, 0.93; in LXe, 0.75) to divide uncorrected $\Phi_i$. Uncertainty in correction factors contributes to the systematic uncertainty in $\Phi_i$ (5\% total)\\ 
\hline
Mirror angle measurements & Place mirror in incident beam, observe viewing angle of reflected beam & Any time sample rack is removed and replaced to swap samples & All, at wavelength of measurements & Determines incident angle dial settings and uncertainty ($\pm2^\circ$) as well as testing for misalignment \\ 
\hline
Mirror power measurements & Observe maximum photon rate of beam reflected off of mirror at several angles, monitor for deviation of more than 5\% & Any time sample rack is removed and replaced to swap samples & All, at 400~nm & Determines correction factor to $\Phi_i$ at high incident angle due to part of beam missing samples \\
\hline
Background subtraction & Scan PMT about chamber with deuterium lamp off, and scan away from the beam with deuterium lamp on and sample rack lifted clear & Each LXe or vacuum run, and if high PMT rate with deuterium lamp off indicates possible light leak & All & Determines rate to subtract from raw data to yield $\Phi_r$ \\
\hline
\end{tabular*}
\end{table*}

\clearpage
\clearpage

\bibliographystyle{JHEP_mod} 
\bibliography{main}   

\end{document}